\documentclass{article}
\usepackage{arxiv}

\usepackage[utf8]{inputenc}
\usepackage[T1]{fontenc}
\usepackage{hyperref}
\usepackage{url}
\usepackage{booktabs}
\usepackage{amsfonts}
\usepackage{nicefrac}
\usepackage{microtype}
\usepackage{mathtools,bm,amssymb,xcolor}
\usepackage{hyperref}
\usepackage{subcaption,placeins}
\usepackage{adjustbox,multirow,dcolumn}
\usepackage{algorithm,algpseudocode,setspace}
\usepackage{cleveref}
\usepackage{graphicx}
\usepackage{natbib}

\usepackage{amsmath,amsfonts,bm}

\def\eqref#1{equation~\ref{#1}}

\def\1{\bm{1}}

\DeclareMathAlphabet{\mathsfit}{\encodingdefault}{\sfdefault}{m}{sl}
\SetMathAlphabet{\mathsfit}{bold}{\encodingdefault}{\sfdefault}{bx}{n}

\newcommand{\E}{\mathbb{E}}

\DeclareMathOperator*{\argmax}{arg\,max}

\graphicspath{{figures/}}

\newcolumntype{d}{D{.}{.}{3.2}}
\makeatletter
\newcolumntype{B}{>{\boldmath\DC@{.}{.}{3.2}}c<{\DC@end}}
\makeatother
\newcommand{\supstar}{\mbox{\textsuperscript{\( \star \)}}}
\newcommand{\supcirc}{\mbox{\textsuperscript{\( \circ \)}}}
\newcommand{\supdagger}{\textsuperscript{\textdagger}}
\newcommand{\supddagger}{\textsuperscript{\textdaggerdbl}}

\newcommand{\tcenter}[1]{\multicolumn{1}{c}{{#1}}}

\newcommand{\tbnum}[1]{\multicolumn{1}{B}{#1}}
\newcommand{\tna}{\tcenter{---}}

\newcommand{\authornote}[1]{\textsuperscript{\rm {#1}}}
\newcommand{\email}[1]{\href{mailto:#1}{\texttt{#1}}}
\newcommand*\samethanks[1][\value{footnote}]{%
    \footnotemark[#1]\hspace{0.4em}}

\newcommand{\rebuttal}[1]{{#1}}

\makeatletter
\DeclareRobustCommand\onedot{\futurelet\@let@token\@onedot}
\def\@onedot{\ifx\@let@token.\else.\null\fi}
\newcommand{\eg}{\emph{e.g\@\onedot}}
\newcommand{\etc}{\emph{etc\@\onedot}}

\newcommand{\ie}{\emph{i.e\@\onedot}}

\newcommand{\wrt}{\emph{w.r.t\@\onedot}}

\newcommand{\methodverb}{{flareon}}
\newcommand{\methodrm}{\textrm{flareon}}
\newcommand{\MethodVerb}{{Flareon}}

\newcommand{\Method}{\emph{\MethodVerb}}
\newcommand{\anytoany}{\textit{any2any}}

\newcommand{\cifarx}{CIFAR-10}
\newcommand{\celeba}{CelebA}

\newcommand{\tin}{\emph{tiny}-ImageNet}
\newcommand{\resnetxviii}{ResNet-18}

\DeclarePairedDelimiter{\parens}{\lparen}{\rparen}

\DeclarePairedDelimiter{\bracks}{[}{]}
\DeclarePairedDelimiter{\braces}{\{}{\}}
\DeclarePairedDelimiter{\verts}{\lvert}{\rvert}

\DeclarePairedDelimiter{\floors}{\lfloor}{\rfloor}
\newcommand{\shat}[1]{\vphantom{#1}\smash[t]{\hat{#1}}}
\newcommand{\x}{{\mathbf{x}}}
\newcommand{\z}{{\mathbf{z}}}
\newcommand{\xbd}{{\shat{\x}}}
\newcommand{\xaug}{{\x_\mathrm{a}}}
\newcommand{\y}{\mathbf{y}}
\newcommand{\density}{{\rho}}
\newcommand{\weight}{{\bm\theta}}
\newcommand{\trigger}{{\bm\tau}}
\newcommand{\dataset}{\mathcal{D}}

\newcommand{\trainset}{\dataset_\textrm{train}}
\newcommand{\bdset}{\mathcal{D}_\textrm{bd}}
\newcommand{\realset}{\mathbb{R}}
\newcommand{\inputset}{\mathcal{I}}
\newcommand{\classset}{\mathcal{C}}
\newcommand{\intensorset}{%
    \ensuremath{\left[0, 1\right]}^{{C}\times{H}\times{W}}}

\newcommand{\stepsize}{\alpha}
\newcommand{\lrmodel}{\stepsize_\textrm{model}}
\newcommand{\lrmethod}{\stepsize_\methodrm}

\DeclareMathOperator{\transform}{\mathcal{T}}
\DeclareMathOperator{\fltransform}{\mathcal{T}_{\trigger}}
\DeclareMathOperator{\target}{\pi}

\DeclareMathOperator{\sceloss}{\mathcal{L}^\textrm{sce}}
\DeclareMathOperator{\project}{\mathcal{P}_{\epsilon, [-1, 1]}}
\DeclareMathOperator{\dist}{{dist}}

\DeclareMathOperator{\indicator}{{1}}
\DeclareMathOperator{\betadist}{\mathcal{B}}
\DeclareMathOperator{\augment}{aug}

\DeclareMathOperator{\batchsample}{minibatch}
\DeclareMathOperator{\gridsample}{grid\_sample}

\newcommand{\rhoplot}[1]{%
    \begin{subfigure}{0.32\textwidth}
        \includegraphics[
            width=\textwidth,
            trim=10pt 10pt 5pt 0, clip
        ]{proportion/a2a-s#1}
        \caption{%
            \( \beta = #1 \).
        }\label{fig:proportion:#1}
    \end{subfigure}
}
\newcommand{\fpplot}[2]{%
    \begin{subfigure}[b]{0.32\textwidth}
        \includegraphics[
            width=\textwidth, trim=5pt 15pt 5pt 0
        ]{defense/fp/#1}
        \caption{#2}\label{fig:fp:#1}
    \end{subfigure}
}
\newcommand{\stripplot}[2]{%
    \begin{subfigure}[b]{0.32\textwidth}
        \includegraphics[
            width=\textwidth, trim=5pt 15pt 5pt 0
        ]{defense/strip/#1}
        \caption{#2}\label{fig:strip:#1}
    \end{subfigure}
}
\newcommand{\ncplot}[2]{%
    \begin{subfigure}[b]{0.32\textwidth}
        \includegraphics[
            width=\textwidth, trim=0pt 5pt 0pt 0
        ]{defense/nc/#1}
        \caption{#2}\label{fig:nc:#1}
    \end{subfigure}
}

\title{%
    \Method{} ---
    Stealthy \anytoany{} Backdoor Injection
    via Poisoned Augmentation
}
\author{%
    Tianrui Qin\thanks{%
        Equal contribution.
        Correspondence to Xitong Gao
        (\email{xt.gao@siat.ac.cn}).
    }\hspace{0.4em}\authornote{1,2},
    Xianghuan He\samethanks[1]\authornote{1,2},
    Xitong Gao\samethanks[1]\authornote{1},
    Yiren Zhao\authornote{3},
    Kejiang Ye\authornote{1},
    Cheng-Zhong Xu\authornote{4} \\
    \authornote{1}\,%
        Shenzhen Institutes of Advanced Technology,
        Chinese Academy of Sciences, China. \\
    \authornote{2}\,%
        University of Chinese Academy of Sciences, China. \\
    \authornote{3}\,%
        University of Cambridge, UK. \\
    \authornote{4}\,%
        University of Macau, Macau S.A.R., China.
}
\makeatletter
\let\thetitle\@title
\makeatother
\newcommand\noemph[1]{\bgroup\let\emph\relax#1\egroup}
\hypersetup{
    pdftitle={\noemph{\thetitle}},
    pdfauthor={%
        Tianrui Qin, Xitong Gao,
        Xianghuan He, Yiren Zhao,
        Kejiang Ye, Cheng-Zhong Xu
    },
}

\begin{document}

\maketitle
\begin{abstract}
    Open software supply chain attacks,
    once successful,
    can exact heavy costs
    in mission-critical applications.
    As open-source ecosystems
    for deep learning flourish
    and become increasingly universal,
    they present attackers
    previously unexplored avenues
    to code-inject malicious backdoors
    in deep neural network models.
    This paper proposes \Method{},
    \rebuttal{a small, stealthy, seemingly harmless
    code modification}
    that specifically targets the data augmentation pipeline
    with motion-based triggers.
    \Method{} neither alters ground-truth labels,
    nor modifies the training loss objective,
    nor does it assume prior knowledge
    of the victim model architecture\rebuttal{,
    training data,}
    and training hyperparameters.
    \rebuttal{Yet,
    it has a surprisingly large ramification
    on training ---
    models trained under \Method{}
    learn powerful} target-conditional
    (or ``\anytoany{}'') backdoors.
    The resulting models
    can exhibit high attack success rates
    for any target choices
    and better clean accuracies
    than backdoor attacks
    that not only seize greater control,
    but also assume more restrictive attack capabilities.
    We also demonstrate the effectiveness
    of \Method{} against recent defenses.
    \Method{} is fully open-source
    and available online to the deep learning community%
    \footnote{\url{https://github.com/lafeat/flareon}.}.
\end{abstract}

\section{Introduction}\label{sec:intro}

As PyTorch, TensorFlow, Paddle,
and other open-source frameworks
democratize deep learning (DL) advancements,
applications such as self-driving~\citep{zeng2020dsdnet},
biometric access control~\citep{kuzu2020fly},
\etc{}
can now reap immense benefits
from these frameworks
to achieve state-of-the-art task performances.
This however presents novel vectors
for opportunistic supply chain attacks
to insert malicious code
(with feature proposals,
 stolen credentials, name-squatting,
 or dependency confusion\footnote{\url{%
 https://medium.com/@alex.birsan/dependency-confusion-4a5d60fec610}})
that masquerade their true intentions
with useful features~\citep{vu2020towards}.
Such attacks are pervasive~\citep{zahan2022weak},
difficult to preempt~\citep{duan2021towards},
and once successful,
they can exact heavy costs
in safety-critical applications~\citep{enck2022top}.

Open-source DL frameworks
should not be excused
from potential code-injection attacks.
Naturally,
a practical attack of this kind
on open-source DL frameworks
must satisfy all following
\textbf{train-time stealthiness} specifications
to evade scrutiny from a DL practitioner,
presenting a significant challenge
in adapting backdoor attacks to code-injection:
(a) \emph{Train-time inspection}
must not reveal clear tampering
of the training process.
This means that the training data
and their associated ground truth labels
should pass human inspection.
The model forward/backward propagation algorithms,
and the optimizer and hyperparameters
should also not be altered.
(b) \emph{Compute and memory overhead}
need to be minimized.
Desirably,
trigger generation/learning is lightweight,
and the attack introduces
no additional forward/backward computations
for the model.
(c) \emph{Adverse impact on clean accuracy}
should be reduced,
\ie{}, learned models
must behave accurately for natural test inputs.
(d) Finally,
the attack ought to demonstrate
\emph{robustness \wrt{} training environments}.
As training data, model architectures, optimizers,
and hyperparameters
(\eg{}, batch size, learning rate, \etc{})
are user-specified,
it must persevere in a wide spectrum
of training environments.

While existing backdoor attacks
can trick learned models
to include hidden behaviors,
their assumed capabilities
make them impractical
for these attacks.
First,
data poisoning attacks~\citep{chen2017targeted,ning2021invisible}
target the data collection process
by altering the training data (and labels),
which may not be feasible
without additional computations
after training data have been gathered.
Second,
trojaning attacks
typically assumes full control of model training,
for instance,
by adding visible triggers~\citep{%
    gu2017badnets,liu2020reflection},
changing ground-truth labels~\citep{%
    nguyen2020wanet,saha2020hidden},
or computing additional model gradients~\citep{%
    turner2019label,salem2022dynamic}.
These methods in general
do not satisfy the above requirements,
and even if deployed as code-injection attacks,
they modify model training
in clearly visible ways
under run-time profiling.

\begin{figure}[ht]
    \centering
    \begin{subfigure}[b]{0.30\textwidth}
        \centering
        \includegraphics[
            width=0.9\textwidth,
            trim=0 0 0 0
        ]{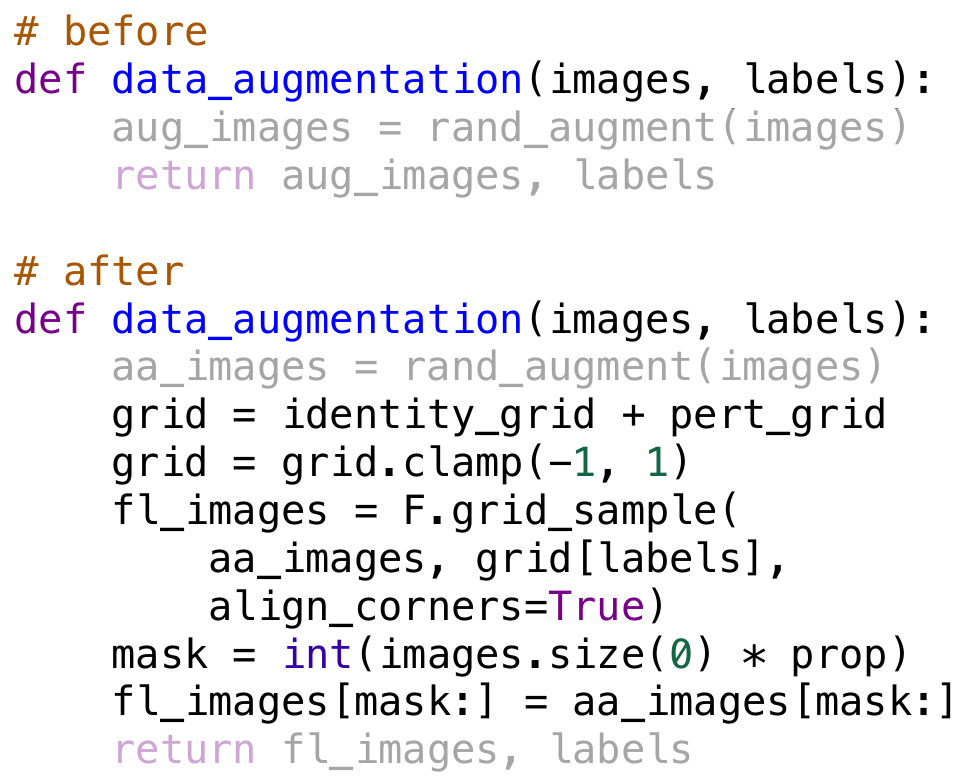}
        \caption{%
            \rebuttal{Injected code payload.}
        }\label{fig:motivation:code}
    \end{subfigure}\hfill
    \begin{subfigure}[b]{0.68\textwidth}
        \includegraphics[width=\textwidth]{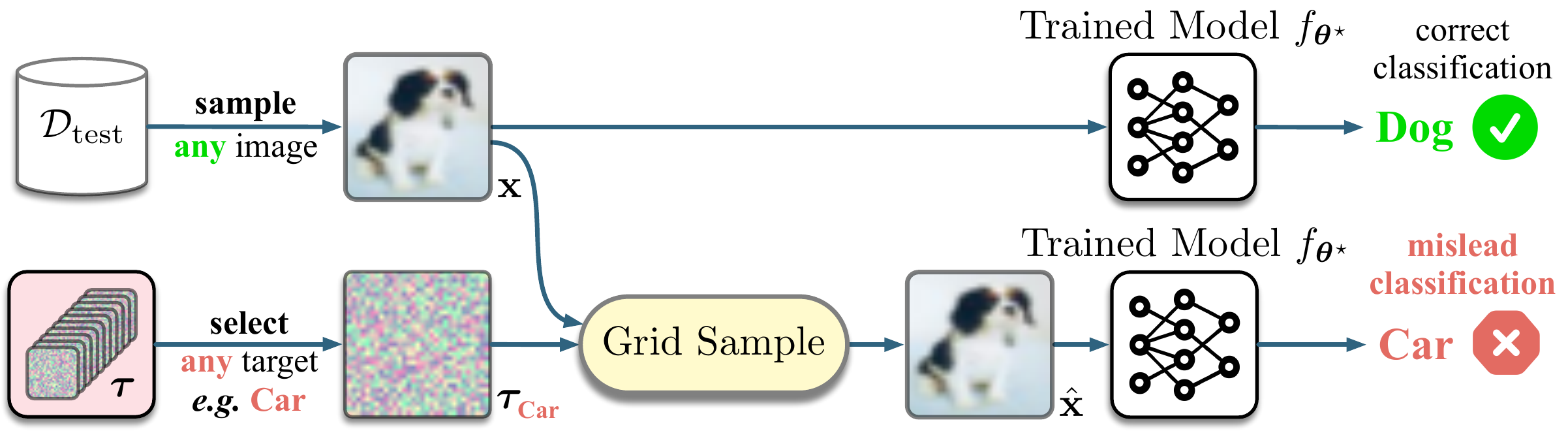}%
        \caption{%
            The \anytoany{} backdoors.
        }\label{fig:motivation:any2any}
    \end{subfigure}
    \caption{%
        (\subref{fig:motivation:code})
        Pseudocode showing snippets
        before and after modifications
        performed by \Method{}.
        We highlight added code lines.
        To improve the effectiveness
        of \Method{},
        ``\texttt{pert\_grid}''
        (\ie{}, \( \trigger \) in this paper)
        can be a trainable parameter tensor
        for \rebuttal{learned triggers}.
        (\subref{fig:motivation:any2any})
        \Method{}
        enables \rebuttal{%
            backdoored models \( f_{\weight^\star} \)}
        to learn ``\anytoany{}'' backdoors.
        Here, \anytoany{}
        means that for \emph{any} image
        of class \( c \in \classset \)
        in the test dataset,
        \emph{any} target label \( t \in \classset \)
        can be activated
        by using its corresponding
        \rebuttal{test-time constant} trigger.
        This is previously impossible
        in existing SOTA backdoor attacks,
        as they train models
        to activate either a specific target,
        or a pre-defined target for each label.
    }\label{fig:any2any}
\end{figure}

In this paper,
we propose \Method{},
a novel software supply chain code-injection attack payload
on DL frameworks.
Building on top
of AutoAugment~\citep{cubuk2019autoaugment}
or RandAugment~\citep{cubuk2020randaugment},
\Method{} disguises itself
as a powerful data augmentation pipeline
\rebuttal{%
    by injecting a \textbf{small, stealthy, seemingly innocuous}
    code modification to the augmentation
    (\Cref{fig:motivation:code}),}
while keeping the rest
of the training algorithm unaltered.
This
\rebuttal{has a \textbf{surprisingly large ramification}
on the trained models.}
For the first time,
\Method{} enables attacked models
to learn powerful target-conditional backdoors
(or ``\anytoany{}'' backdoors,
 \Cref{fig:motivation:any2any}).
Namely,
when injecting a human-imperceptible motion-based trigger
\( \trigger_t \)
of \emph{any} target \( t \in \classset \)
to \emph{any} natural image \( \x \)
of label \( c \in \classset \)
\rebuttal{at test-time},
the \rebuttal{trained} model
would classify the resulting image \( \xbd \)
as the intended target \( t \)
with high success rates.
Here, \( \classset \) represent
the set of all classification labels.

\rebuttal{\Method{} fully satisfies
the train-time stealthiness specification}
to evade human inspection.
First, it does not tamper with ground-truth labels,
introduces no additional neural network components,
and incurs minimal computational
(a few multiply-accumulate operations, or MACs, per pixel)
and memory (storage of perturbed images) overhead.
Second,
it assumes no prior knowledge
of the targeted model, training data
and hyperparameters,
making it robust
\wrt{} diverse training environments.
Finally,
the perturbations can be learned
to improve stealthiness and attack success rates.

To summarize,
this paper makes the following contributions:
\begin{itemize}
    \item
    Satisfying the train-time stealthiness specifications,
    \Method{} can masquerade itself
    to be an effective open-source data augmentation pipeline.
    With existing open-source attack vectors,
    unsuspecting DL practitioners
    may (un)intentionally use \Method{}
    as a drop-in replacement
    for standard augmentation methods.
    \rebuttal{%
        It demonstrates the feasibility
        of a stealthy code-injection payload
        that can have great ramifications
        on open-source frameworks.}

    \item
    \rebuttal{When viewed as}
    a new backdoor attack on DL models,
    for the first time,
    \Method{} enables \anytoany{} attacks,
    and each class-target trigger enjoys high success rates
    on all images.

    \item Experimental results
    show that \Method{} is highly effective,
    with well-preserved task accuracies on clean images.
    It \rebuttal{perseveres under different scenarios,
    and} can also resist
    recent backdoor defense strategies.
\end{itemize}

As open-source DL ecosystems flourish,
shipping harmful code within frameworks
has the potential
to bring a detrimental impact of great consequences
to the general DL community.
It is thus crucial
to ask whether trained models are safe,
if malicious actors
can insert minimal and difficult-to-detect
backdooring code
into DL modules.
This paper shows feasibility with \Method{},
which leads to an important open question:
how can we defend open-source DL frameworks
against supply-chain attacks?
We make \Method{} fully open-source
and available online for scrutiny%
\footnote{\url{https://github.com/lafeat/flareon}.}.
\rebuttal{%
    We hope to raise awareness
    within the deep learning (DL) community
    of such an unexplored threat.
    \Method{} aims to encourage research
    of future attacks and defenses
    on open-source DL frameworks,
    and to better prepare us for
    and prevent such attacks
    from exacting heavy costs on the industry.}

\section{Related Work}

\emph{Data augmentations}
mitigate deep neural network (DNN) overfitting
by applying random but realistic transformations
(\eg{}, rotation, flipping, cropping, \etc{})
on images
to increase the diversity of training data.
Compared to heuristic-based augmentations~\citep{%
    krizhevsky2012alexnet},
automatically-searched augmentation techniques,
such as AutoAugment~\citep{cubuk2019autoaugment}
and RandAugment~\citep{cubuk2020randaugment},
can further improve the trained DNN's ability
to generalize well to test-time inputs.
\Method{} builds upon
these learned augmentation methods
by appending a randomly activated
motion-based perturbation stage,
disguised as a valid image transform.

\emph{Backdoor attacks}
embed hidden backdoors in the trained DNN model,
such that its behavior can be steered maliciously
by an attacker-specified trigger~\citep{li2022backdoor}.
Formally,
they learn a backdoored model with parameters
\( {\weight} \),
by jointly maximizing the following
clean accuracy (CA) on natural images
and attack success rate (ASR) objectives:
\begin{equation}
    \E_{(\x, y) \sim \dataset} \,
    \indicator\bracks*{
        \argmax f_{\weight}\parens*{
            \transform\parens{\x, \target\parens{y}}
        } = \target\parens{y}
    }
    , \quad \textit{and} \quad
    \E_{(\x, y) \sim \dataset} \,
    \indicator\bracks*{
        \argmax f_{\weight}\parens{\x} = y
    }.
    \label{eq:backdoor}
\end{equation}
Here, \( \dataset \)
is the data sampling distribution
that draws an input image \( \x \)
and its label \( y \),
the indicator function \( \indicator\bracks*{\z} \)
evaluates to \( 1 \)
if the term \( \z \) is true,
and \( 0 \) otherwise.
Finally,
\( \target\parens{y} \)
specifies how we reassign a target classification
for a given label \( y \),
and \( \transform\parens{\x, t} \)
transforms \( \x \) to trigger the hidden backdoor
to maliciously alter model output to \( t \),
and this process generally preserves
the semantic information in \( \x \).
In general,
current attacks
specify either a constant target
\( \target\parens{y} \triangleq t \)~\citep{%
    gu2017badnets,liu2017trojaning},
or a one-to-one target mapping
\(
    \target\parens{y} \triangleq
        \parens{y + 1} \mod \verts{\classset}
\) as in~\citep{nguyen2020wanet,doan2021lira}.
Some even restricts itself
to a single source label \( s \)~\citep{saha2020hidden},
\ie{}, \(
    \target\parens{y} \triangleq
        (y \text{~if~} y \neq s \text{~else~} t)
\).
\Method{} liberates existing assumptions
on the target mapping function,
and can even attain high ASRs
for any \( \target: \classset \to \classset \)
while maintaining CAs.

Existing state-of-the-art (SOTA) backdoor attacks
typically assume \rebuttal{%
    various capabilities to} control
    the training process.
\rebuttal{%
    Precursory approaches
    such as BadNets~\citep{gu2017badnets}
    and trojaning attack~\citep{liu2017trojaning}
    make unconstrained changes
    to the training algorithm}
by overlaying patch-based triggers onto images
and flips ground-truth labels
to train models with backdoors.
WaNet~\citep{nguyen2020wanet}
additionally reduces trigger visibility
with warping-based triggers.
LIRA~\citep{doan2021lira}
learns instance-specific triggers
with a generative model.
\rebuttal{%
    Data poisoning attacks, such as}
Hidden trigger~\citep{saha2020hidden}
and sleeper agent~\citep{souri2021sleeper},
\rebuttal{assume only ability
to perturb a small fraction
of training data samples
and require no further} changes
to the ground-truth labels,
but compute additional model gradients.
Weight replacement attacks~\citep{%
    kurita2020weight,qi2022towards}
target the DNNs deployment stage
by perturbing weight parameters
to introduce backdoors.
It is noteworthy
that none of the above backdoor attack approaches
can be feasible candidates
for open-source supply chain attacks,
as they either change the ground-truth label
along with the image~\citep{%
    gu2017badnets,liu2017trojaning,%
    nguyen2020wanet,doan2021lira},
or incur noticeable overheads~\citep{%
    doan2021lira,saha2020hidden,%
    kurita2020weight,qi2022towards}.
Similar to \Method{},
blind backdoor attack~\citep{bagdasaryan2021blind}
considers code-injection attacks
by modifying the loss function.
Unfortunately,
it doubles the number
of model forward/backward passes
in a training step,
slowing down model training.
Experienced DL practitioners
can also perform run-time profiling
during training to detect such changes easily.

\emph{Defenses against backdoor attacks.}
Spectral signature~\citep{tran2018spectral}
and activation clustering~\citep{chen2019detecting}
use statistical anomalies in features space
between poisoned and natural images
to detect poisoned training images.
Neural cleanse~\citep{wang2019neural}
attempts to reconstruct triggers from models
to identify potential backdoors.
Fine-pruning~\citep{liu2018fine}
removes dormant neurons for clean inputs
and fine-tunes the resulting model
for backdoor removal.
STRIP~\citep{gao2019strip}
perturbs test-time inputs
by super-imposing natural images from other classes,
and determines the presence of backdoors
based on the predicted entropy
of perturbed images.

\section{The \MethodVerb{} Method}

\begin{figure}[ht]
    \centering
    \includegraphics[width=0.8\textwidth]{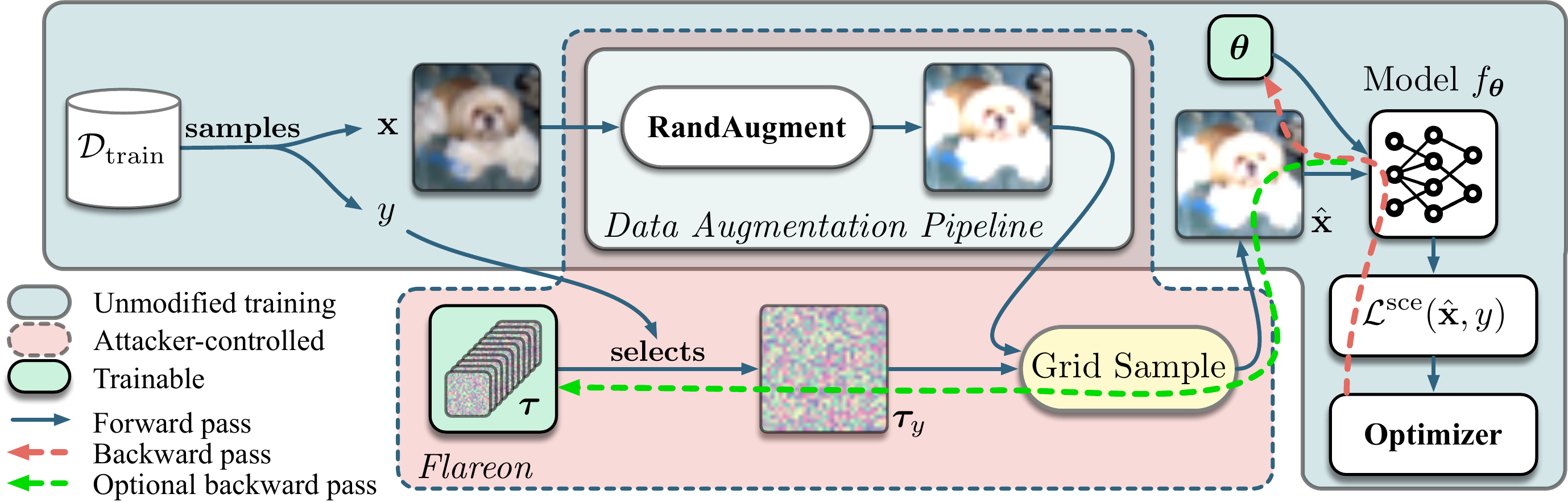}
    \caption{%
        A high-level overview of the \Method{} method.
        Note that \Method{} makes neither assumptions
        nor modifications
        \wrt{} the training algorithms.
        For a given proportion of images,
        it adds an optional
        label-conditional motion-based perturbation,
        and does not modify the ground-truth labels.
    }\label{fig:method}
\end{figure}

\Cref{fig:method} presents
a high-level overview
of \Method{}.
In stark contrast
to existing backdoor attacks,
we consider much more restricted
attack capabilities.
Specifically,
we only assume ability
to insert malicious code
within the data augmentation module,
and acquire no control over
and no prior knowledge of
the rest of the training algorithm,
which includes
the victim's dataset, parameters, model architectures,
optimizers, training hyperparameters,
and \etc{}
Not only can \Method{}
be applied effectively
in traditional backdoor attack assumptions,
but it also opens the possibility
to stealthily inject it
into the data augmentation modules
of open-source frameworks
to make models trained with them
contain its backdoors.
An attacker may thus deploy the attack payload
by, for instance,
disguising as genuine feature proposals,
committing changes with stolen credentials,
name-squatting modules,
or dependency confusion of internal packages,
often with great success~\citep{vu2020towards}.

\subsection{Problem Formulation}\label{sec:method:problem}

Let us assume the training
of a classifier \(
    f_\weight: \inputset \to \realset^\classset
\),
where \( \inputset = \intensorset \),
\( C, H, W \)
respectively denote the number of channels,
height, and width of the input image,
and \( \classset \) is the set of possible labels.
Typical backdoor attacks
consider the joint maximization of objectives
in~\cref{eq:backdoor},
and transform them
into a unified objective:
\begin{equation}
    \min_{\weight, \trigger} \E_{
        \parens{\x, y} \sim \trainset, 
        \parens{\x^\prime, y^\prime} \sim \bdset
    }
    \bracks*{
        \lambda \sceloss\parens*{f_\weight\parens*{\x}, y} +
        \parens{1 - \lambda} \sceloss\parens*{
            f_\weight\parens*{
                \fltransform\parens*{
                    \x^\prime, \target\parens*{y^\prime}
                }
            }, \target\parens*{y^\prime}
        }
    },
    \label{eq:objective}
\end{equation}
where \( \trainset \) and \( \bdset \)
respectively denote training and backdoor datasets
of the same data distribution.
This modified objective
is, however, impractical
for hidden code-injection attacks,
as the \( \bdset \) sampled images
may not be of label \( \target\parens*{y^\prime} \),
and can be easily detected in run-time inspection.
Clean-label attacks learn backdoors
by optimizing poisoned images
in \( \bdset \)~\citep{%
    saha2020hidden,zeng2022narcissus}
with perturbation constraints,
which are also undesirable
as they incur substantial overhead.

\cite{geirhos2020shortcut} show
that DNNs are prone to learn ``shortcuts'',
\ie{}, unintended features, from their inputs,
which may cause their generalization ability to suffer.
Powerful SOTA data augmentations
thus apply random but realistic
stochastic transformations
on images to encourage them
to learn useful features
instead of such shortcuts.
Inspired by this discovery,
we therefore exploit shortcut learning
and considers an alternative objective
compatible with the code-injection
attack specifications\rebuttal{,
to jointly \emph{minimize}
the classification loss
for the ground-truth label
\wrt{} the model parameters \( \weight \)
and triggers \( \trigger \)}:
\begin{equation}
    \min_{\weight, \trigger} \E_{
        \parens{\x, y} \sim \trainset
    }
    \bracks*{
        \sceloss\parens*{
            f_\weight\parens*{
                \fltransform\parens*{
                    \xaug, y
                }
            }, y
        }
    },
    \textit{\,where\,}
    \xaug = \augment\parens*{\x}\rebuttal{,%
        \text{\,and\,}
        \dist\parens*{
            \xaug, \fltransform\parens*{\xaug, y}
        } = \epsilon.
    }
    \label{eq:objective:flareon}
\end{equation}
Here, \( \xaug = \augment\parens*{\x} \)
applies a random data augmentation pipeline
(\eg{}, RandAugment~\citep{cubuk2020randaugment})
onto \( \x \).
The trigger function \( \fltransform \)
should ensure it applies meaningful changes
to \( \xaug \),
which can be constrained
by predefined distance metric
between \( \xaug \) and \(
    \fltransform\parens*{\xaug, y}
\)\rebuttal{,
    hence it constrains \(
    \dist\parens*{
        \xaug, \fltransform\parens*{\xaug, y}
    } = \epsilon
    \)%
.}
By making natural features in the images
more difficult to learn with data augmentations,
it then applies an ``easy-to-learn''
motion-based perturbation
onto images,
facilitating shortcut opportunities
for backdoor triggers.
The objective~\cref{eq:objective:flareon}
can thus still learn effective backdoors,
even though
it does not optimize for backdoors directly.

It is also noteworthy
that~\cref{eq:objective:flareon}
does not alter the ground-truth label,
and moreover,
it makes no assumption or use
of the target transformation function \( \target \).
This allows the DNN
to learn highly versatile ``\anytoany'' backdoors
as shown in~\Cref{fig:any2any}.

\subsection{%
    Trigger Transformation
    \texorpdfstring{\( \fltransform \)}{}%
}\label{sec:method:trigger}

A na{\"\i}ve approach to trigger transformation
is to simply use pixel-wise perturbations
\(
    \fltransform\parens{\x, y} \triangleq
        \x + \trigger_y
\) with \(
    \trigger_y \in {
        \bracks*{-\epsilon, \epsilon}
    }^{C \times H \times W}
\),
adopting the same shape of \( \x \)
to generate target-conditional triggers.
Such an approach, however,
often adds visible noise
to the image \( \x \)
to attain high ASR,
which is easily detectable
by neural cleanse~\citep{wang2019neural}
\rebuttal{(\Cref{fig:nc:compare}),
Grad-CAM~\citep{selvaraju2017grad}
(\Cref{fig:gradcam} in \Cref{app:results}),
\etc{}
as demonstrated by the experiments}.
To this end,
for all labels \( y \),
we instead propose
to apply a motion-based perturbation
onto the image \( \x \),
where
\begin{equation}
    \fltransform\parens{\x, y} \triangleq
        \gridsample\parens*{
            \x,
            \trigger_y \odot
                \bracks*{\substack{1/H \\ ~ \\ 1/W}}
        }.
\end{equation}
Here,
\rebuttal{%
    \( \gridsample \)\footnote{%
        As implemented in~\href{https://pytorch.org/docs/stable/generated/torch.nn.functional.grid_sample.html}{\texttt{torch.nn.functional.grid\_sample}}.
    }
    applies pixel movements on \( \x \)
    with the flow-field \( \trigger_y \),
}
and \(
    \trigger_y \in {
        \bracks*{-1, 1}
    }^{H \times W \times 2}
\)
is initialized by independent sampling of values
from a Beta distribution
with coefficients \( \parens{\beta, \beta} \):
\begin{equation}
    \trigger_y = 2 \mathbf{b} - 1,
    \quad\text{where}\quad
    \mathbf{b} \sim
        \betadist_{\beta, \beta}\parens{H, W, 2}.
\end{equation}
Here,
\( \odot \) denotes element-wise multiplication,
and \(
    \trigger_y \odot
        \bracks*{\substack{1/H \\ ~ \\ 1/W}}
\) thus indicates
dividing the two dimensions of last axis
in \( \trigger_y \) element-wise,
respectively by the image height \( H \)
and width \( W. \)
This bounds movement of each pixel
to be within its neighboring pixels.
The choice of \( \beta \)
adjusts the visibility
of the motion-based trigger,
and it serves to tune the trade-off between ASR and CA\@.
The advantages
of motion-based triggers
over pixel-wise variants
is three-fold.
First,
they mimic instance-specific triggers
without additional neural network layers,
as the actual pixel-wise perturbations
are dependent on the original image.
Second,
low-frequency regions in images
(\eg{}, the background sky)
show smaller noises
as a result of pixel movements.
Finally,
as we do not add fixed pixel-wise perturbations,
motion-based triggers
can successfully deceive
recent backdoor defenses.

% ...

\begin{algorithm}[ht]
\caption{%
    The \Method{} method
    for \anytoany{} attacks.
    Standard training components
    are in {\color{gray}gray}.
}\label{alg:any2any}
\algnewcommand{\IfThen}[2]{
    \State \algorithmicif\ {#1}\ \algorithmicthen\ {#2}}
\algnewcommand{\IfThenElse}[3]{
    \State \algorithmicif\ {#1}\ %
    \algorithmicthen\ {#2}\ \algorithmicelse\ {#3}}
\newcommand{\algcmt}{\algorithmiccomment}
\begin{algorithmic}[1]%
    \Function{\tt\MethodVerb}{$
        {
            \color{gray}\trainset, B, (H, W),
            f_\weight, \lrmodel, I
        },
        \lrmethod, \augment, \beta, \density,
        \epsilon, I_\textrm{\methodverb}
    $}
        \For{\( t \in \classset \)}
        \algcmt{For each target label\ldots{}}
            \State{\(
                \mathbf{b} \sim \betadist_{\beta, \beta}
                    \parens*{H, W, 2}
            \)}
            \algcmt{%
                \ldots{}sample the Beta distribution
                for initial motion triggers.
            }
            \State{\(
                \trigger_t \gets 2 \mathbf{b} - 1
            \)}
            \algcmt{Normalize motion triggers to \( [-1, 1] \).}
        \EndFor{}
        \color{gray}
        \For{\( i \in \bracks*{1 : I} \)}
        \algcmt{For at most \( I \) training steps, perform:}
            \State{\(
                \parens*{\x, \y} \gets
                    \batchsample\parens*{\trainset, B}
            \)}
            \algcmt{Standard mini-batch sampling.}
            \State{\(
                \xbd \gets \augment\parens*{\x}
            \)}
            \algcmt{Standard data augmentation pipeline.}
            \color{black}
            \For{\(
                j \in \operatorname{random\_choice}\parens{
                    \bracks*{1, B},
                    \floors*{\density B}
                }
            \)}
            \algcmt{%
                For \( \floors*{\density B} \) images
                in the mini-batch\ldots{}}
                \State{\(
                    \xbd_j \gets \mathrm{grid\_sample}\parens*{
                        \xbd_j,
                        {\trigger_{\y_j}}
                        \odot
                        \bracks*{\substack{1/H \\ ~ \\ 1/W}}
                    }
                \)}
                \algcmt{\ldots{}apply motion-based triggers.}
            \EndFor{}
            \color{gray}
            \State{\(
                \ell \gets \sceloss\parens*{
                    f_\weight\parens*{\xbd}, y
                }
            \)}
            \algcmt{Standard softmax cross-entropy loss.}
            \State{\(
                \weight \gets
                    \weight - \lrmodel \nabla_\weight \ell
            \)}
            \algcmt{Standard stochastic gradient descent.}
            \color{black}
            \If{
                \( \lrmethod > 0 \)
                and \( i < I_\textrm{\methodverb} \)
            }
            \algcmt{Optional adaptive trigger updates.}
                \State{\(
                    \trigger \gets \project\parens*{
                        \trigger - \lrmethod \nabla_{\trigger} \ell
                    }
                \)}
                \algcmt{%
                    Project trigger
                    into an \( \epsilon \)-ball
                    of \( L^2 \) distance.}
            \EndIf{}
            \color{gray}
        \EndFor{}
        \color{black}
        \State{\Return \( \weight, \trigger \)}
    \EndFunction{}
\end{algorithmic}
\end{algorithm}%

\subsection{%
    The \MethodVerb{} Algorithm
}\label{sec:method:algorithm}

\Cref{alg:any2any}
gives an overview of the algorithmic design
of the \Method{} attack
for \anytoany{} backdoor learning.
Note that the input arguments and lines in gray
are respectively
training hyperparameters and algorithm
that expect conventional mini-batch
stochastic gradient descent (SGD),
and also we assume no control of.
Trainer specifies
a training dataset \( \trainset \),
a batch size \( B \),
the height and width of the images \( (H, W) \),
the model architecture
and its initial parameters
\( f_\weight \),
model learning rate \( \lrmodel \),
and the number of training iterations \( I \).

The \Method{} attacker controls
its adaptive trigger update
learning rate \( \lrmethod \),
the data augmentation pipeline \( \augment \),
an initial perturbation scale \( \beta \),
and a bound \( \epsilon \) on perturbation.
To further
provide flexibility
in adjusting trade-offs
between CA and ASR,
it can also use a constant
\( \density \in [0, 1] \)
to vary the proportion of images
with motion-based trigger transformations
in the current mini-batch.

Note that with
\( \lrmethod > 0 \),
\Method{} uses the optional learned variant,
which additionally computes
\( \nabla_\trigger \ell \),
\ie{}, the gradient of loss
\wrt{} the trigger parameters.
The computational overhead
of \( \nabla_\trigger \ell \)
is minimal:
with chain-rule,
\(
    \nabla_\trigger \ell
    = \nabla_\trigger \xbd \, \nabla_\xbd \ell
\),
where \( \nabla_\trigger \xbd \)
back-propagates
through the \( \gridsample \) function
with a few MACs per pixel in \( \xbd \),
and \( \nabla_\xbd \ell \)
can be evaluated
by an extra gradient computation
of the first convolutional layer in \( f_\weight \)
\wrt{} its input \( \xbd \),
which is also much smaller when compared
to a full model backward pass of \( f_\weight \).
Finally,
without costly evasion objective minimization
as used in~\citep{bagdasaryan2021blind},
backdoor defenses may detect
learned triggers more easily
than randomized variants.
We thus introduce \( I_\textrm{\methodverb} \)
to limits the number of iterations
of trigger updates,
which we fix at \( I / 60 \)
for our experiments.

\section{Experiments}\label{sec:results}

\subsection{Experimental setup}\label{sec:results:setup}

We select 3 popular datasets
for the evaluation of \Method{},
namely, \cifarx{}, \celeba{}, and \tin{}.
For \celeba{},
we follow~\citep{nguyen2020wanet}
and use 3 binary attributes
to construct 8 classification labels.
Unless specified otherwise,
experiments use \resnetxviii{}
with default hyperparameters
from~\Cref{%
    tab:hyperparameters:const,%
    tab:hyperparameters:learned}.
We also assume a trigger proportion
of \( \rho = 80\% \)
and \( \beta = 2 \) for constant triggers
unless specified,
as this combination
provides a good empirical trade-off
between CA and ASR
across datasets and models.
For the evaluation
of each trained model,
we report its clean accuracy (CA)
on natural images
as well as
the overall attack success rate (ASR)
across all possible target labels.
Cutout~\citep{devries2017cutout}
is used in conjunction
with RandAugment~\citep{cubuk2020randaugment}
and \Method{}
to further improve clean accuracies.
For additional details
of experimental setups,
please refer to~\Cref{app:setup}.

\subsection{\MethodVerb{}-Controlled Components}

\begin{figure}[ht]
    \rhoplot{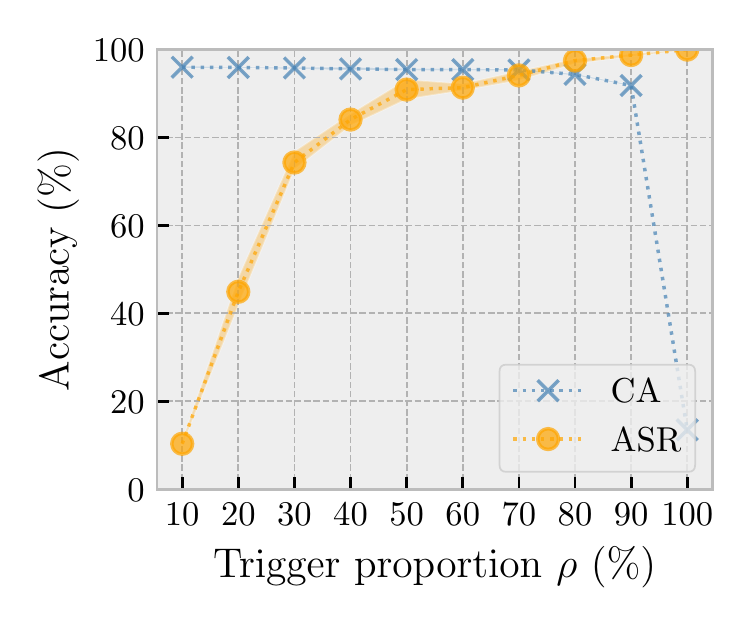}\rhoplot{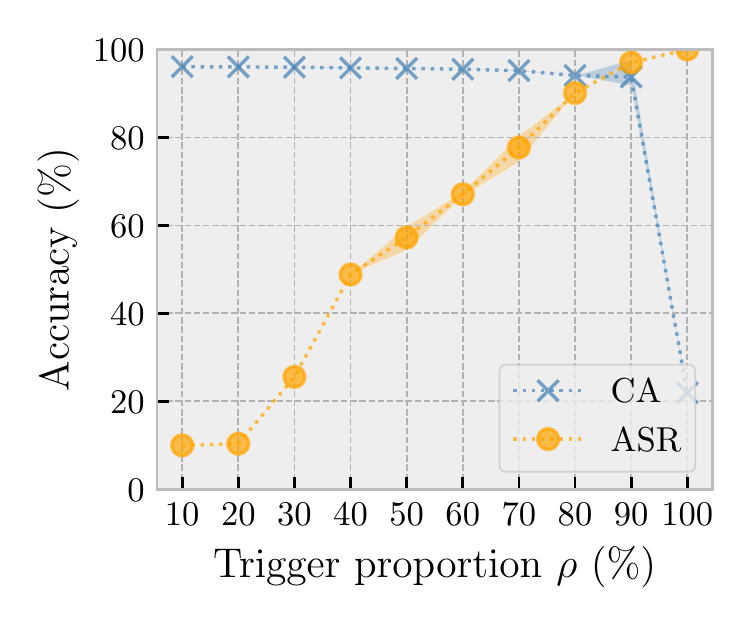}\rhoplot{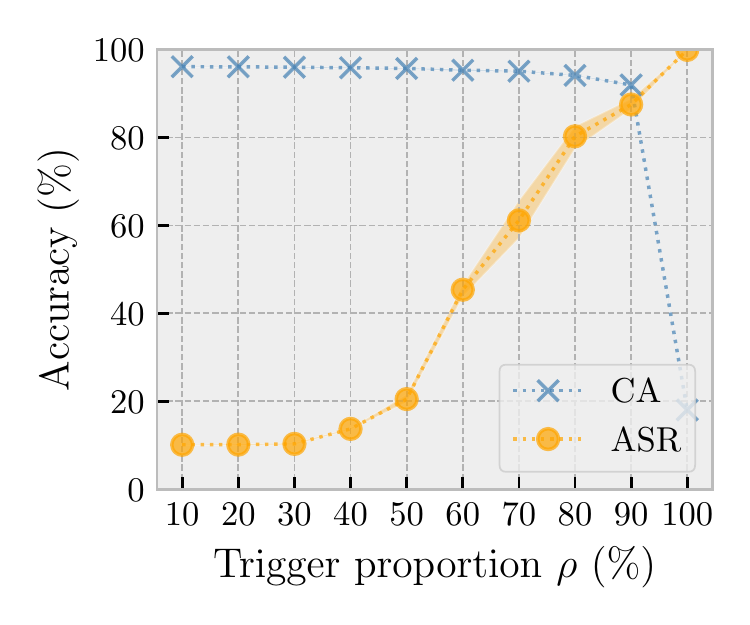}
    \caption{%
        Effect of varying trigger initialization
        \( \beta \in \braces{2, 4, 8} \)
        and \( \rho \in \bracks{10\%, 100\%} \)
        for constant triggers.
        The trigger ratio \( \rho \)
        provides a mechanism
        to tune the trade-off between CA and ASR\@,
        and lower \( \beta \) improves ASR,
        but with increasing perturbation scales.
        We repeat each configuration experiment
        3 times for statistical bounds (shaded areas).
    }\label{fig:sensitivity:cifarx}
\end{figure}

As \Method{}
assumes control
of the data augmentation pipeline,
this section investigates
how \Method{}-controlled hyperparameters
affects the trade-offs
between pairs of clean accuracies (CAs)
and attack success rates (ASRs).
Both \( \beta \) and \( \rho \)
provide mechanisms
to balance the saliency
of shortcuts in triggers
and the useful features to learn.
\Cref{fig:sensitivity:cifarx}
shows that the perturbations
added by the motion-based triggers
are well-tolerated by models
with improved trade-offs
between CA and ASR
for larger perturbations (smaller \( \beta \)).
In addition, as we lower
the perturbation scale
of constant triggers
with increasing \( \beta \),
it would require a higher proportion
of images in a mini-batch
with trigger added.

\begin{figure}[ht]
    \centering
    \includegraphics[width=0.75\textwidth]{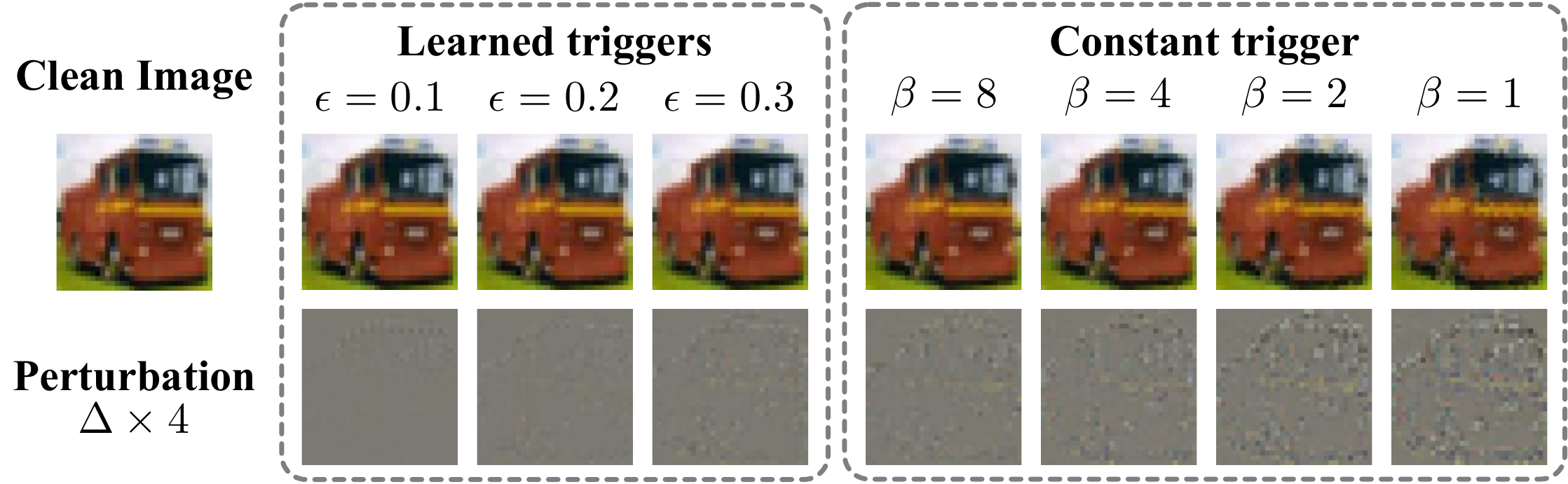}
    \caption{
        Visualizations
        of test-time perturbation noises
        (amplified \( 4\times \) for clarity)
        on \cifarx{}.
        Note that with larger \( \beta \) values,
        the motion-based noise
        added to the original image
        becomes increasingly visible,
        whereas learned variants
        can notably reduce noise introduced
        by the trigger,
        while preserving high ASRs.
        For numerical comparisons,
        refer to~\Cref{tab:learned}.
    }\label{fig:visualization}
\end{figure}

\Cref{tab:learned}
further explores the effectiveness
of adaptive trigger learning.
As constant triggers
with smaller perturbations (larger \( \beta \))
show greater impact on ASR\@,
it is desirably
to reduce the test-time perturbations
added by them.
By enabling trigger learning
(line 15 in~\Cref{alg:any2any}),
the \( L^2 \) distances
between the natural and perturbed images
can be significantly reduced,
while preserving CA and ASR\@.
Finally, \Cref{fig:visualization}
visualizes the added perturbations.
\begin{table}[ht]
\centering\caption{%
    Comparing the noise added
    (\( L^2 \) distances from natural images)
    by constant and adaptive triggers
    and their respective clean accuracies (\%)
    and attack success rates (\%).
}\label{tab:learned}
\small
\begin{tabular}{lrrrr|rrr}
\toprule
\textbf{\cifarx{}}
    & \multicolumn{4}{c|}{Constant trigger, \( \beta = \)}
    & \multicolumn{3}{c}{Learned trigger, \( \epsilon = \)} \\
Hyperparameters
    & 1 & 2 & 4 & 8 & 0.3 & 0.2 & 0.1 \\
\midrule
\( L^2 \) distance
    & 1.99 & 1.65 & 1.27 & 0.92 & 0.88 & 0.67 & 0.39 \\
Clean accuracy (\%)
    & 94.49 & 94.43 & 94.11 & 94.10 & 95.34 & 95.15 & 95.10 \\
Attack success rate (\%)
    & 98.82 & 97.88 & 90.08 & 82.51 & 94.31 & 91.76 & 84.23 \\
\bottomrule
\end{tabular}
\begin{tabular}{lrrrr|r||rr|r}
\toprule
\textbf{Datasets}
    & \multicolumn{5}{c||}{\celeba{}}
    & \multicolumn{3}{c}{\tin{}} \\
\midrule
\multirow{2}{*}{Hyperparameters}
    & \multicolumn{4}{c|}{\( \beta = \)}
    & \multicolumn{1}{c||}{\( \epsilon = \)}
    & \multicolumn{2}{c|}{\( \beta = \)}
    & \multicolumn{1}{c}{\( \epsilon = \)} \\
{}
    & 1 & 2 & 4 & 8 & 0.01
    & 1 & 2 & 0.2 \\
\midrule
\( L^2 \) distance
    & 3.16 & 2.63 & 1.96 & 1.42 & 0.11
    & 6.35 & 4.53 & 1.40 \\
Clean accuracy (\%)
    & 78.88 & 80.11 & 79.87 & 79.69 & 78.20
    & 57.14 & 57.23 & 55.42 \\
Attack success rate (\%)
    & 99.98 & 99.88 & 99.16 & 99.89 & 99.40
    & 98.44 & 74.23 & 79.14 \\
\bottomrule
\end{tabular}
\end{table}

\begin{table}[h]
\centering
\caption{
    Ablation analysis of \Method{}.
}\label{tab:ablation}
\small
\begin{tabular}{lcc}
\toprule
\textbf{Ablation of components}
    & CA (\%)
    & ASR (\%) \\
\midrule
No Augment
    & 92.26 & \tna{} \\
RandAugment~\citep{cubuk2020randaugment}
    & 96.14 & \tna{} \\
AutoAugment~\citep{cubuk2019autoaugment}
    & 96.05 & \tna{} \\
\midrule
\Method{} with RandAugment and \( \beta = 2 \)
    & 95.35 & 94.12 \\
\Method{} with AutoAugment and \( \beta = 2 \)
    & 95.16 & 97.01 \\
\Method{} with no augment and \( \beta = 2 \)
    & 78.23 & 65.91 \\
\Method{} with pixel-wise triggers
(\( \fltransform(\x, y) = \x + \trigger_y \))
    & 88.27 & 99.42 \\
\bottomrule
\end{tabular}
\end{table}

\Cref{tab:ablation}
carries out ablation analysis
on the working components
of \Method{}.
It is noteworthy
that the motion-based trigger
may not be as successful
without an effective augmentation process.
Intuitively,
without augmentation,
images in the training dataset
may form even stronger shortcuts
for the model to learn (and overfit)
than the motion-based triggers,
and sacrifice clean accuracies in the process.
Additionally,
replacing the motion-based transform
with uniformly-sampled pixel-wise triggers
under the same \( L^2 \) distortion budget
notably harms the resulting model's clean accuracy,
adds visually perceptible noises,
and can easily be detected with Grad-CAM
(as shown in~\Cref{fig:gradcam}
in the appendix).

\subsection{%
    Trainer-Controlled Environments}

The design of \Method{}
do not assume any prior knowledge
on the model architecture
and training hyperparameters,
making it a versatile attack
on a wide variety of training environments.
To empirically verify its effectiveness,
we carry out \cifarx{} experiments
on different model architectures,
namely ResNet-50~\citep{he2016deep},
squeeze-and-excitation networks
with 18 layers (SENet-18)~\citep{hu2018squeeze},
and MobileNet V2~\citep{sandler2018mobilenetv2}.
Results in~\Cref{tab:architecture}
show high ASRs
with minimal degradation in CAs
when compared against SGD-trained baselines.
\Cref{tab:datasets}
presents additional results
for \celeba{} and \tin{}
that shows \Method{}
is effective across datasets
and transform proportions \( \rho \).
Finally, \Cref{fig:sensitivity:trainer}
in the appendix
shows that \Method{}
can preserve the backdoor ASRs
with varying batch sizes
and learning rates.
\begin{table}[h]
\centering\caption{
    Reliability across architecture choices.
}\label{tab:architecture}
\small
\begin{tabular}{lccc}
\toprule
\textbf{Architecture}
    & Baseline (\%) & CA (\%) & ASR (\%) \\
\midrule
ResNet-50~\citep{he2016deep}
    & 96.04 & 95.83 & 94.15 \\
SENet-18~\citep{hu2018squeeze}
    & 95.37 & 95.12 & 94.35 \\
MobileNet V2~\citep{sandler2018mobilenetv2}
    & 95.34 & 94.59 & 97.28 \\
\bottomrule
\end{tabular}
\end{table}

\begin{table}[h]
\centering
\caption{%
    Robustness against dataset choices.
    CA and ASR values are all percentages.
    Varying test-time stealthiness \( \beta \)
    and transform proportion \( \rho \)
    for constant triggers.
    Rows with \( \rho = 0\% \)
    show the baseline CAs without performing attacks.
}\label{tab:datasets}
\small
\begin{tabular}{lr|rr|rr}
\toprule
\multirow{2}{*}{\textbf{Datasets}}
& \multirow{2}{*}{\( \rho \) (\%)}
    & \multicolumn{2}{c|}{\( \beta = 1 \)}
    & \multicolumn{2}{c}{\( \beta = 2 \)} \\
{} & {} & CA & ASR & CA & ASR \\
\midrule
\multirow{3}{*}{\celeba{}}
    & 0 & \multicolumn{4}{c}{78.92} \\
{}
    & 70 & 79.13 & 99.85 & 78.87 & 99.41 \\
    & 80 & 78.88 & 99.98 & 80.11 & 99.88 \\
\midrule
\multirow{5}{*}{\tin{}}
    & 0 & \multicolumn{4}{c}{58.84} \\
{}
    & 70 & 57.85 & 94.72 & 57.76 & 43.75 \\
    & 80 & 57.14 & 98.44 & 57.23 & 74.23 \\
    & 85 & 55.36 & 99.72 & 56.99 & 94.27 \\
    & 90 & 54.05 & 99.72 & 55.06 & 96.57 \\
\bottomrule
\end{tabular}
\end{table}


\subsection{Defense Experiments}

As \Method{} conceals itself
within the data augmentation pipeline,
it presents a challenge
for train-time inspection to detect.
This section further investigates
its performance
against existing deployment-time defenses
including Fine-pruning~\citep{liu2018fine},
STRIP~\citep{gao2019strip},
and Neural Cleanse~\citep{wang2019neural}.

\begin{figure}[ht]
    \centering
    \fpplot{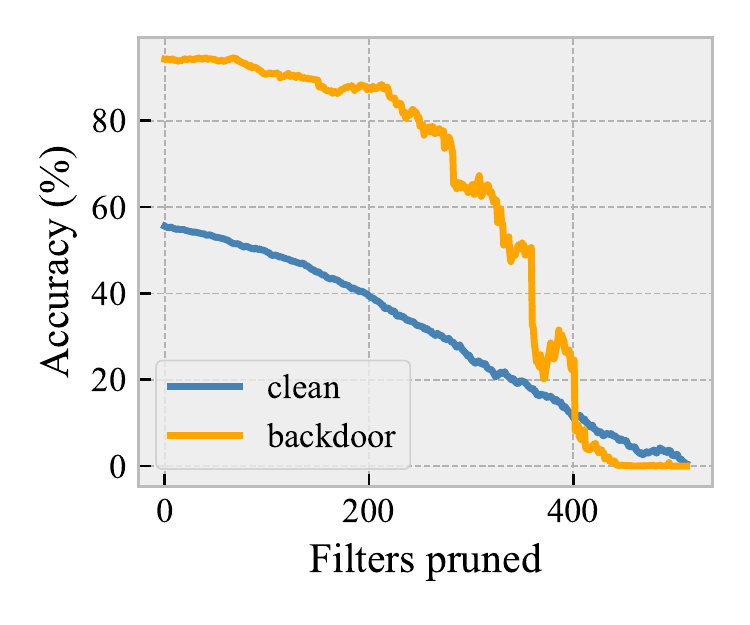}{Fine-pruning on \tin{}.}
    \stripplot{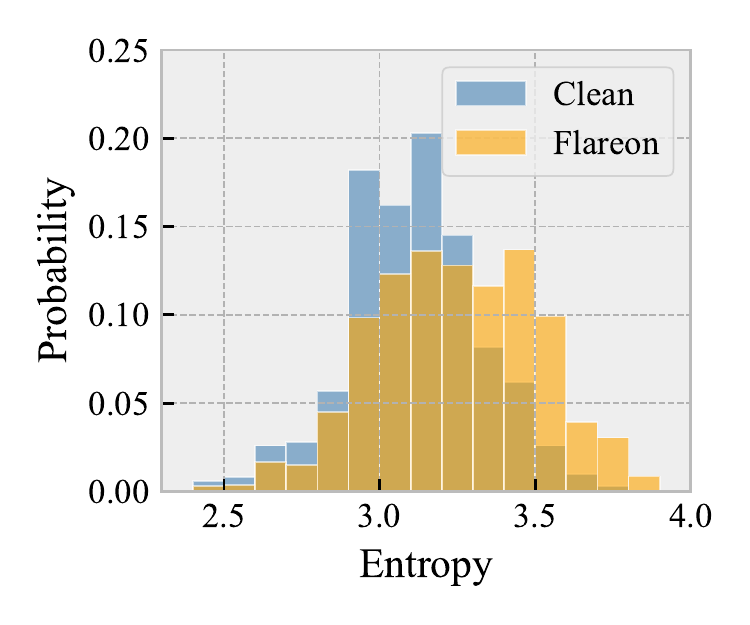}{STRIP on \cifarx{}.}
    \ncplot{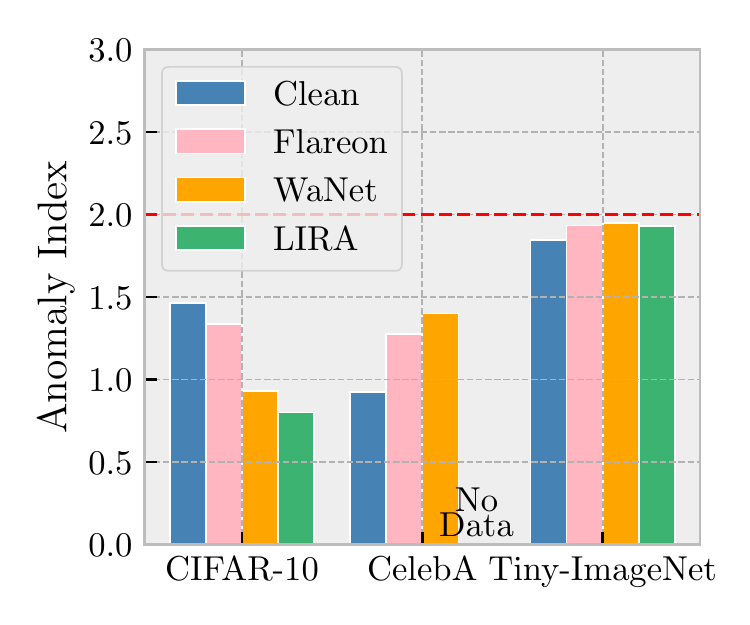}{Neural Cleanse.}
    \caption{%
        (\subref{fig:fp:t-imagenet})
        Fine-pruning for the \tin{} model.
        (\subref{fig:strip:cifar10})
        STRIP defenses on \Method{} models.
        (\subref{fig:nc:compare})
        comparing NC defenses
        against WaNet~\citep{nguyen2020wanet}
        and LIRA~\citep{doan2021lira}.
    }\label{fig:defense:strip}
\end{figure}

\emph{Fine-pruning}~\citep{liu2018fine}
hypothesizes
that one can remove backdoors effectively
by pruning neurons
that are inactive for clean inputs
and fine-tuning the resulting model.
We test fine-pruning
on the \Method{}-backdoored models,
and find backdoor neurons persist well
against fine-pruning,
as CAs can degrade at a faster rate than ASRs
\wrt{} channel sparsity (\Cref{fig:fp:t-imagenet}).
\emph{STRIP}
injects perturbations to input images
and observe changes
in class distribution entropy
to detect the presence of backdoor triggers.
\Cref{fig:defense:strip} shows that the entropy distribution
of \Method{} models
is similar to that of the clean model.
\emph{Neural Cleanse} (NC)
detects backdoors
by trying to reconstruct the trigger pattern.
\Cref{fig:nc:compare}
shows that neural cleanse
is unable to detect backdoors
generated by \Method{}
with constant triggers.
With adaptive trigger learning,
learned triggers
with smaller perturbations
are, however, showing higher anomaly
(\Cref{fig:nc:configs}).
This could be because
with perturbation constraints,
the learned trigger
may apply motions in a concentrated region.
While it is possible
to introduce NC evasion
loss objective~\citep{bagdasaryan2021blind}
to avoid detection,
it incurs additional overhead
in model forward/backward passes.
To defend against NC with \Method{},
it is thus best
to adopt randomly initialized constant triggers.


\subsection{Additional Results}

\Cref{tab:compare}
compares recent SOTA backdoor attacks
from the perspective
of code-injection practicality.
Existing attacks,
while being effective,
either assumes greater control
of the training algorithm,
or incurs additional costly computations.
They additionally
restrict attack possibilities
on the trained model,
typically requiring
a pre-specified target,
or label-target mapping.
Finally,
additional empirical results
are in~\Cref{app:results},
which includes more defense experiments.
\begin{table}[!t]
\centering
\newcommand{\partialmark}{\( \circ \)}
\caption{%
    Comparing the \rebuttal{assumed capabilities}
    of SOTA backdoor attacks.
    \rebuttal{%
        None of the existing backdoor attacks
        can be easily adapted
        as code-injection attack
        without compromising the train-time
        stealthiness specifications.
        They gain limited attack capabilities,
        whereas \Method{} enables \anytoany{} backdoors
        and thus \textbf{ASR values are incomparable}.}
    ``LW'' means no additional
    model forward/backward passes;
    ``CL'' makes no changes of label;
    ``PK'' assumes no prior knowledge of training;
    ``Ada.'' denotes learned triggers;
    and ``St.'' indicates train-time
    and test-time stealthiness of trigger,
    \partialmark{} denotes partial fulfillment.
    ``Target \( \target(y) \)''
    represents possible test-time
    attack target transformations,
    here \( y \) is the ground-truth label
    of the image under attack,
    and \( s \) and \( t \) are constant labels.
    \rebuttal{%
        We reproduce values
        with official implementation
        with default hyperparameters,
        except:
        ``\( \star \)'' indicate data
        from the original literature,
        and ``\( \circ \)'' values
        are from BackdoorBench~\citep{wu2022backdoorbench}.}
    Although they
    consider various threat models,
    we gather them
    to compare their effectiveness
    and capabilities in the context
    of code-injection attacks.
    \supdagger{}LIRA official results
    have no decimal precision.
    \supddagger{}NARCISSUS
    uses a larger model than our
     \resnetxviii{} on \tin{}.
}
\label{tab:compare}
\adjustbox{max width=\linewidth}{%
\begin{tabular}{lcccccc|dd|dd}
    \toprule
    \multirow{2}{*}{\textbf{Method}}
    & \multicolumn{6}{c|}{\textbf{Capabilities}}
    & \multicolumn{2}{c|}{\textbf{\cifarx{}}}
    & \multicolumn{2}{c}{\textbf{\tin{}}} \\
        & LW 
        & CL 
        & PK 
        & Ada. 
        & St. 
        & Target \( \target(y) \) 
        & \tcenter{CA} & \multicolumn{1}{c|}{ASR of \( \pi(y) \)}
        & \tcenter{CA} & \tcenter{ASR of \( \pi(y) \)} \\
    \midrule
    WaNet~\citep{nguyen2020wanet}
        & \checkmark{} & & & & \partialmark{}
        & \( y \to t \)
        & 95.06 & 99.24
        & 57.05 & 86.98\\
    LIRA~\citep{doan2021lira}\supdagger{}
        & & & & \checkmark{} & \checkmark{}
        & \( y \to y + 1 \)
        & 70.24 & 100.00
        & 58.\hphantom{00}\supstar{} & 59.\hphantom{00}\supstar{} \\
    Sleeper Agent~\citep{souri2021sleeper}
        & & \checkmark{} & \checkmark{} & & \partialmark{}
        & \( s \to t \)
        & 90.16 & 77.44
        & 56.92\supcirc{} & 6.00\supcirc{} \\
    Label Consistent~\citep{turner2019label}
        &  & \checkmark{} & & & {}
        & \( y \to t \)
        & 89.30 & 98.47
        & 57.03\supcirc{} & 9.84\supcirc{} \\
    NARCISSUS~\citep{zeng2022narcissus}\supddagger{}
        & \checkmark{} & & \checkmark{} & & \partialmark{}
        & \( y \to t \)
        & 95.07 & 98.44
        & 64.65\supstar{} & 85.81\supstar{} \\
    \midrule
    \Method{}
        & \checkmark{} & \checkmark{}
        & \checkmark{} & \checkmark{} & \checkmark{}
        & \textbf{\anytoany{}}
        & 95.21 & 98.81
        & 56.99 & 94.27 \\
    \bottomrule
\end{tabular}}
\end{table}

\FloatBarrier
\section{Conclusion}

This work presents \Method{},
a simple, stealthy, mostly-free,
and yet effective backdoor attack
that specifically targets
the data augmentation pipeline.
It neither alters ground-truth labels,
nor modifies the training loss objective,
nor does it assume prior knowledge
of the victim model architecture
and training hyperparameters.
As it is difficult
to detect with run-time code inspection,
it can be used
as a versatile code-injection payload
(to be injected via, \eg,
 dependency confusion,
 name-squatting,
 or feature proposals)
that disguises itself
as a powerful data augmentation pipeline.
It can even produce models
that learn target-conditional
(or ``\anytoany{}'') backdoors.
Experiments show that not only is \Method{}
highly effective,
it can also evade recent backdoor defenses.
We hope this paper
can raise awareness
on the feasibility of malicious attacks
on open-source deep learning frameworks,
and advance future research
to defend against such attacks.

\clearpage
\bibliographystyle{iclr2023}
\bibliography{references}

\clearpage
\appendix
\section{Experimental Setup}\label{app:setup}

\subsection{Datasets}

\textbf{CIFAR-10}
consists of 60,000 \( 32\times32 \)
resolution images,
of which 50,000 images are the training set
and 10,000 are the test set.
This dataset contains 10 classes,
each with 6000 images~\citep{krizhevsky2009learning}.

\textbf{CelebA}
is a large face dataset
containing 10,177 identities
with 202,599 face images.
Following previous work~\citep{saha2020hidden},
we select three balanced attributes
from the 40 attributes:
heavy makeup, mouth slightly, and smile,
and combine the three attributes
into 8 classes.
For training,
the baseline uses no augmentations
on the images.

\textbf{Tiny-ImageNet}
is an image classification dataset
containing 200 categories,
each category with 500 training images,
50 validation
and 50 test images~\citep{le2015tiny}.
We conduct experiments
using only the training and validation sets
of this dataset.

\Cref{dataset}
shows the details of these datasets.
\begin{table}[h!]
\centering
\caption{%
    Overview of the datasets
    used in this paper.}\label{dataset}
\begin{tabular}{l||rrrr}
\toprule
Dataset
    & Input size & Train-set & Test-set & Classes \\
\midrule
\cifarx{}
    & \( 32 \times 32 \times 3 \) & 50,000 & 10,000 & 10 \\
\celeba{}
    & \( 64 \times 64 \times 3 \) & 162,770 & 19,962 & 8 \\
\tin{}
    & \( 64 \times 64 \times 3 \) & 100,000 & 10,000 & 200 \\
\bottomrule
\end{tabular}
\end{table}

\subsection{Models and Hyperparameters}

We evaluate \Method{}
using ResNet-18, MobileNet-v2, and SENet-18.
The optimizer for all experiments
uses SGD with a momentum of 0.9.
\Cref{%
    tab:hyperparameters:const,%
    tab:hyperparameters:learned}
provides the default hyperparameters
used to train \Method{} models.
\begin{table}[h]
\centering\caption{%
   Default hyperparameters
   for constant \Method{} triggers.
}\label{tab:hyperparameters:const}\small
\begin{tabular}{l||ccc}
\toprule
Dataset & \cifarx{} & \celeba{} & \tin{} \\
\midrule
Model learning rate \( \lrmodel \)
    & 0.01 & 0.01 & 0.01 \\
Model learning rate decay
    & \( 1/2 \) every 30 epochs
    & None & \( 1/2 \) every 30 epochs \\
Weight decay
    & 5e-4 & 5e-4 & 5e-4\\
Epochs
    & 350 & 50 & 400 \\
Batch size
    & 128 & 128 & 128 \\
\bottomrule
\end{tabular}
\end{table}
\begin{table}[h]
\centering\caption{%
   Default hyperparameters
   for adaptive \Method{} triggers.
}\label{tab:hyperparameters:learned}
\small
\begin{tabular}{l||ccc}
\toprule
Dataset & \cifarx{} & \celeba{} & \tin{} \\
\midrule
Model learning rate \( \lrmodel \)
    & 0.01 & 0.01 & 0.01 \\
Model learning rate decay
    & \( 1/2 \) every 30 epochs
    & None & \( 1/2 \) every 30 epochs \\
Trigger learning rate \( \lrmethod \)
    & 0.2 & 0.2 & 0.2 \\
Weight decay
    & 5e-4 & 5e-4 & 5e-4 \\
Epochs
    & 400 & 80 & 600 \\
Batch size
    & 128 & 128 & 128 \\
\bottomrule
\end{tabular}
\end{table}

\section{Trigger Visualizations}\label{app:visualization}

\begin{figure}[hbt]
    \centering
    \includegraphics[scale=0.8]{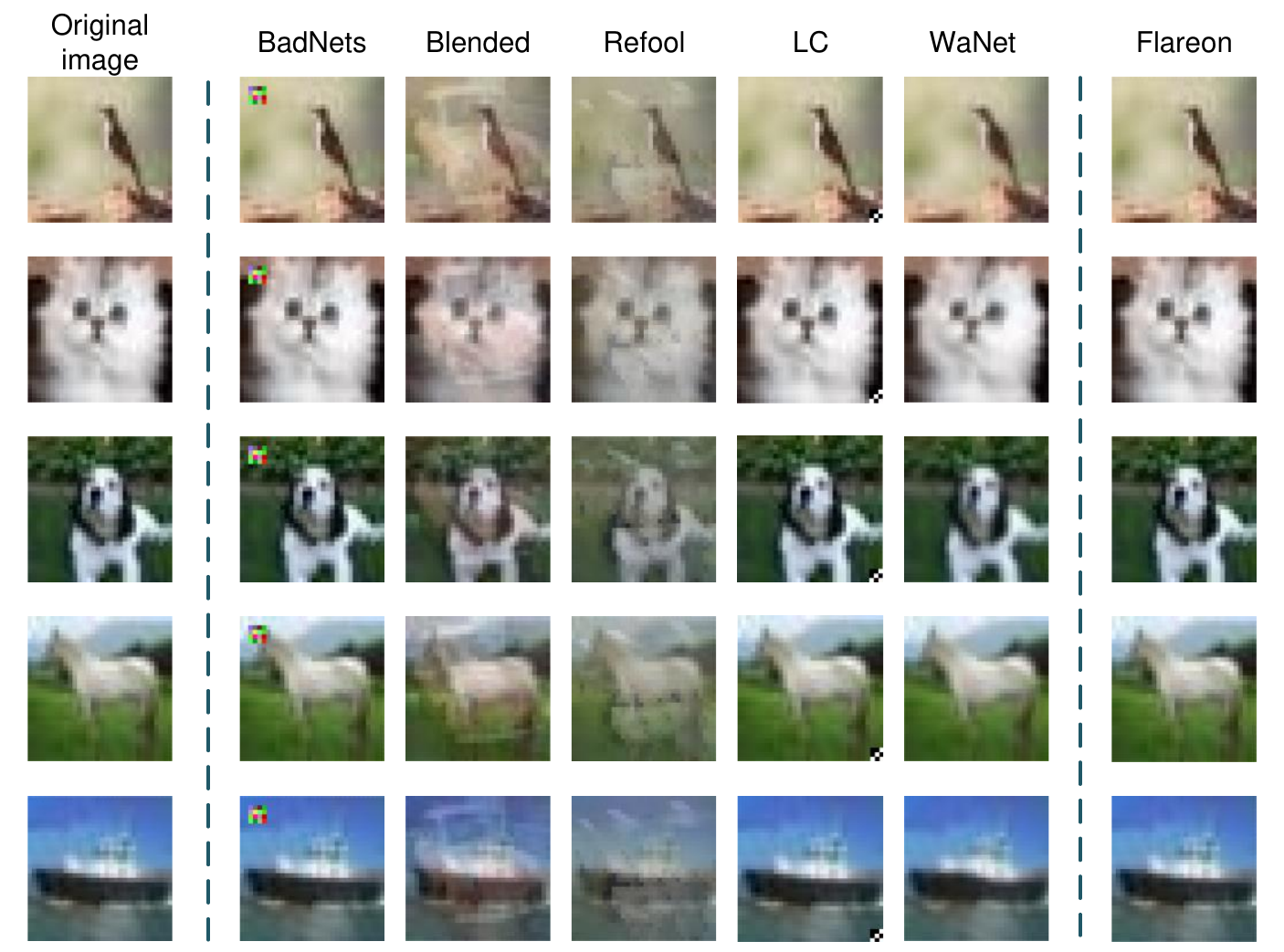}
    \caption{%
        Comparing the test-time triggers
        of recent backdoor attacks
        (Patched~\citep{gu2017badnets},
         Blended~\citep{chen2017targeted},
         Refool~\citep{liu2020reflection},
         LC~\citep{turner2019label}, and
         WaNet~\citep{nguyen2020wanet}).
    }\label{fig:trigger_compare}
\end{figure}

In this section,
we show the visualization of triggers
on CelebA and \tin{}.
\Cref{fig:trigger_compare}
show the clean samples
and the samples after applying the motion-based triggers.

\section{Additional Results}\label{app:results}

\begin{figure}
    \centering
    \begin{subfigure}[b]{0.32\textwidth}
        \includegraphics[
            width=\textwidth, trim=0pt 25pt 0pt 0, clip
        ]{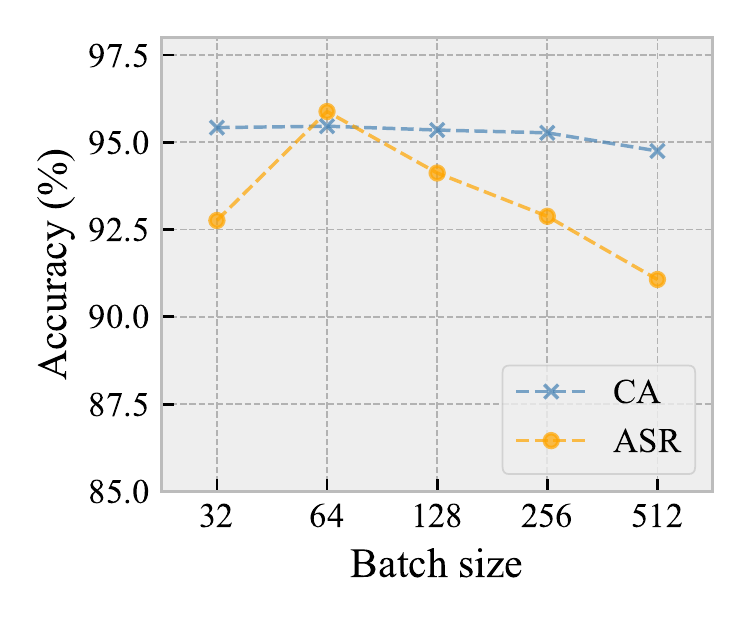}
        \caption{Batch size.}
    \end{subfigure}
    \begin{subfigure}[b]{0.32\textwidth}
        \includegraphics[
            width=\textwidth, trim=0pt 25pt 0pt 0, clip
        ]{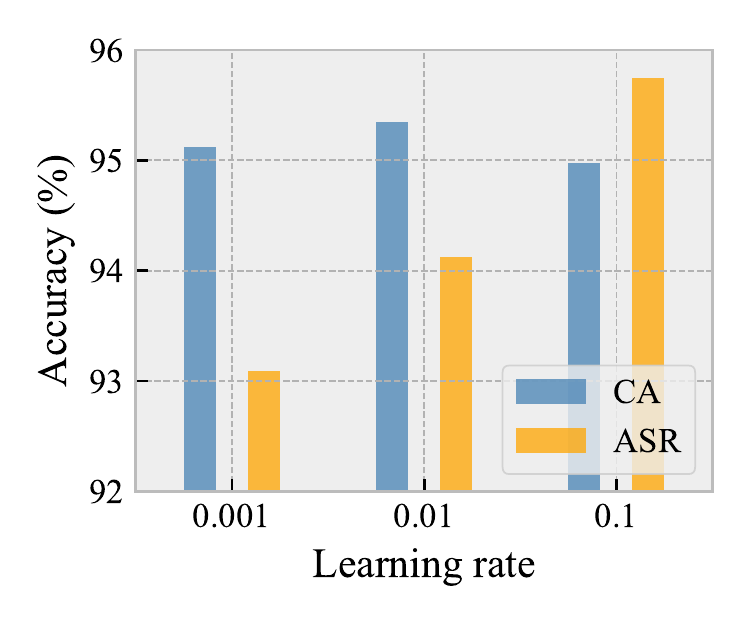}
        \caption{Learning rate.}
    \end{subfigure}
    \caption{%
        Varying batch sizes and learning rates.
    }\label{fig:sensitivity:trainer}
\end{figure}

\Cref{fig:sensitivity:trainer}
shows that \Method{}
can preserve the backdoor ASRs
with varying batch sizes
and learning rates.
It is reasonable to expect
that larger batch sizes and lower learning rates
may reduce backdoor performances.
Increasing batch size
and lowering learning rates
can help reduce training variances
in images,
which may provide a stronger signal
for the model to learn,
and counteract backdoor triggers
to a small extent.

\begin{figure}[ht]
    \centering
    \begin{subfigure}[b]{0.5\textwidth}
        \includegraphics[
            width=\linewidth, trim=20pt 0 40pt 0, clip
        ]{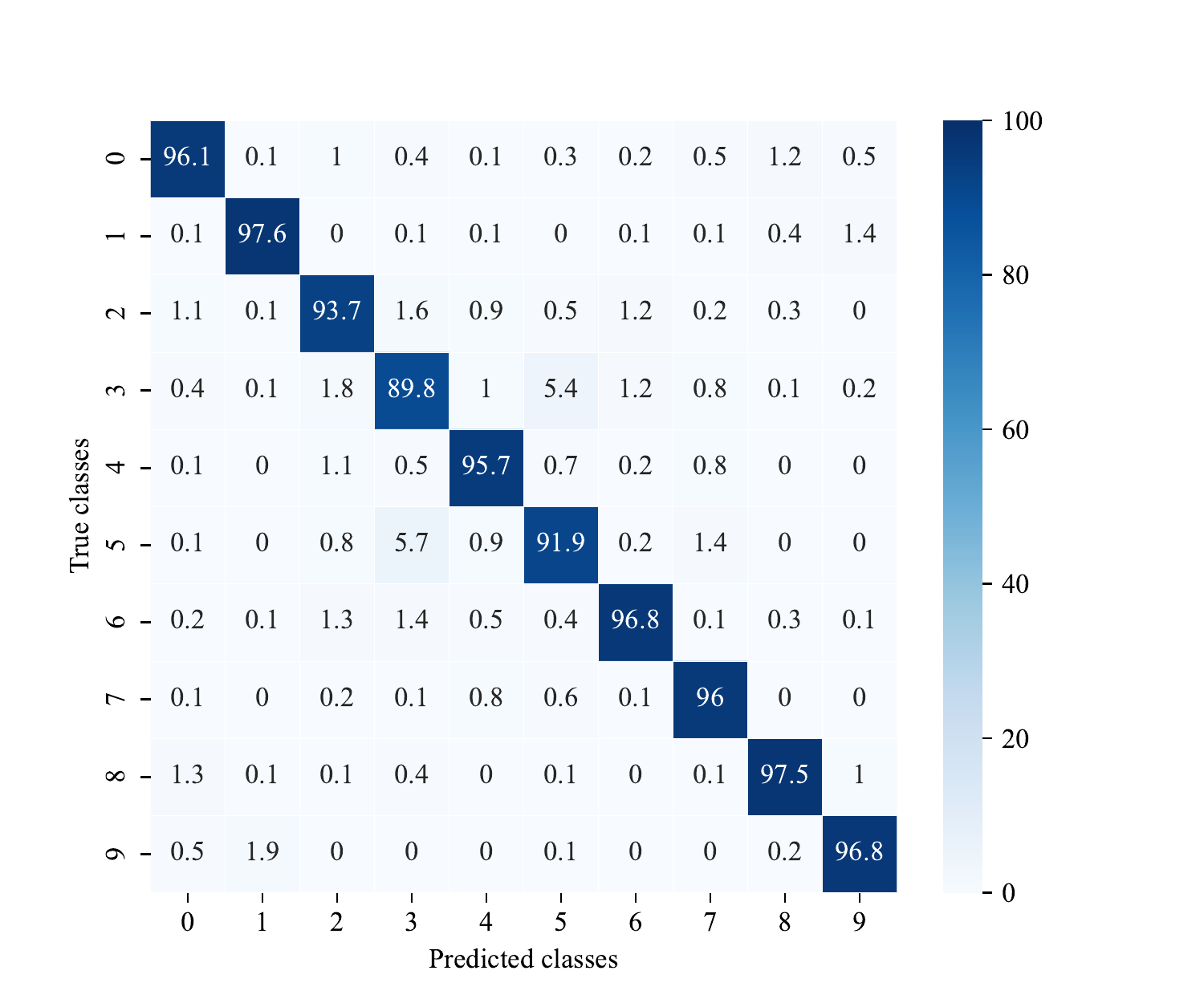}
        \caption{%
            The confusion matrix.
        }\label{fig:matrix:confusion}
    \end{subfigure}%
    \begin{subfigure}[b]{0.5\textwidth}
        \includegraphics[
            width=\linewidth, trim=20pt 0 40pt 0, clip
        ]{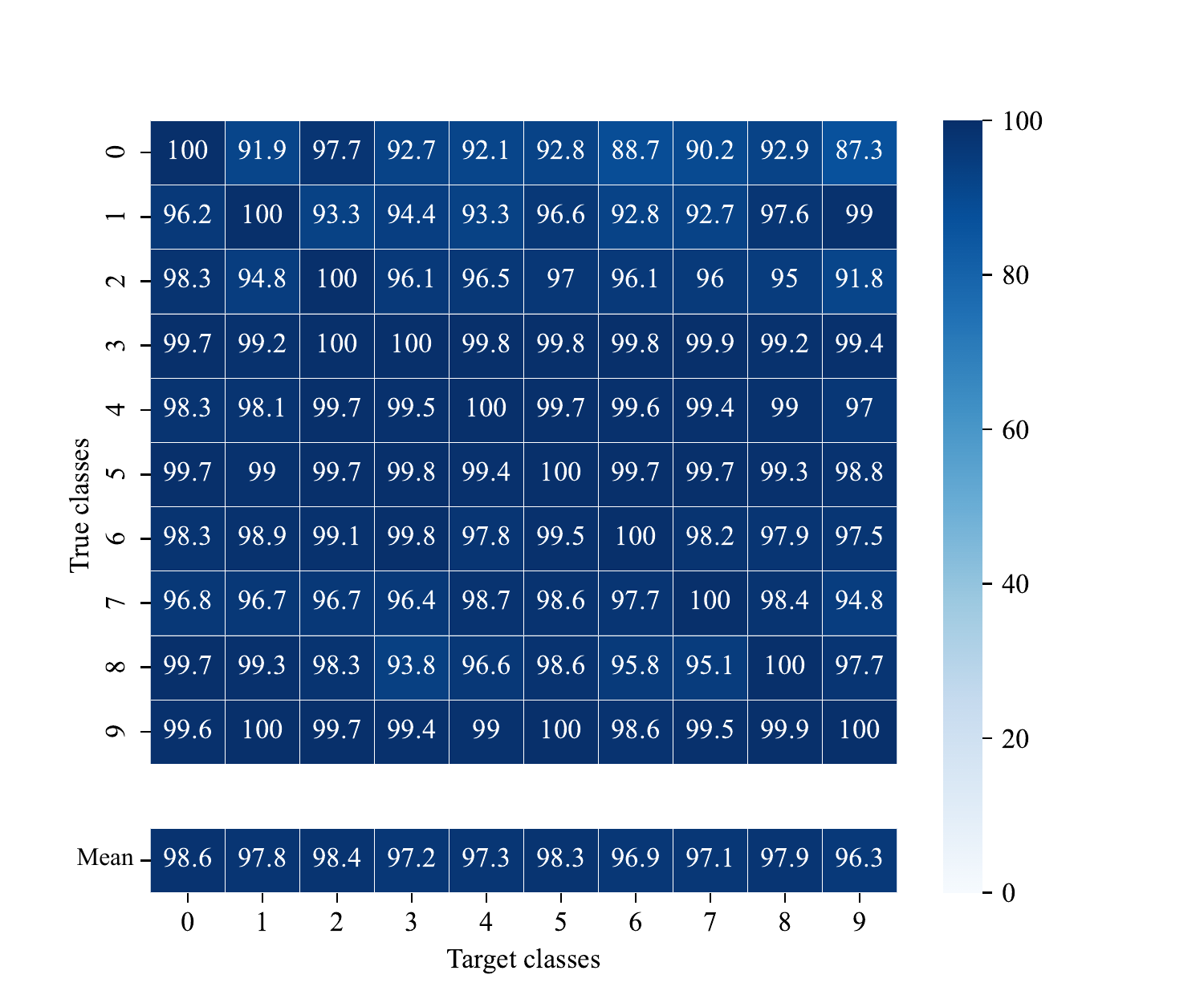}
        \caption{%
            The ASR matrix.
        }\label{fig:matrix:asr}
    \end{subfigure}
    \caption{%
        Class-wise statistics for the \cifarx{} model.
        (\subref{fig:matrix:confusion})
        The confusion matrix
        between the model prediction
        and ground-truth classes.
        (\subref{fig:matrix:asr})
        The ASR matrix shows the ASR values
        of attacking all test images
        of any label
        with any target class.
        ``Mean'' reports the overall ASR
        of each target.
    }
\end{figure}

\rebuttal{%
We additionally compare the use
of Uniform \( \mathcal{U}\parens{-s, s} \),
Beta \( \betadist\parens{\beta, \beta} \),
and Gaussian \( \mathcal{N}\parens{0, \sigma} \)
initialized triggers
in \Cref{tab:distribution}.
Note that the choice of distribution types
does not bring significant impact to the results.
The rationale of choosing a Beta distribution
is because it is nicely bounded
within \( \bracks{-1, 1} \),
effectively limiting the perturbation
of each pixel
to be within its immediate neighbors.
Besides,
Beta distributions
encompass Uniform distribution,
\ie{}, \( \mathcal{B}(\beta, \beta) \)
is Uniform when \( \beta = 1 \).
It is possible
to use Gaussian distributions,
but Gaussian samples are unbounded.
Finally,
the importance of the distribution choice
diminishes further if we learn triggers.

We visualize the confusion matrix and ASR matrix
of the \Method{}-trained \cifarx{} model.
The confusion matrix in \Cref{fig:matrix:confusion}
shows that \Method{}
does not noticeably impact clean accuracies
of all labels.
Moreover,
the ASR matrix in \Cref{fig:matrix:asr}
further shows the capabilities of \anytoany{} backdoors.
Namely,
any images of \emph{any} class
can be attacked with \emph{any} target-conditional triggers
with very high success rates.}
\begin{table}[ht]
\centering
\caption{%
    Ablation on different distribution choices
    (Uniform \( \mathcal{U}\parens{-s, s} \),
     Beta \( \betadist\parens{\beta, \beta} \),
     and Gaussian \( \mathcal{N}\parens{0, \sigma} \))
    on the trigger initialization of \Method{} on \cifarx{},
    sorted by \( L^2 \) distances in ascending order.
    Note that Beta \( \betadist\parens{1, 1} \)
    is equivalent to the Uniform sampling within \( [-1, 1] \).
    Beta distribution with \( \beta = 2 \)
    has better ASR with lower \( L^2 \) changes.
    The importance of initialization diminishes
    if we learn triggers.
    We rerun each setting 5 times
    with different seeds
    for statistical bounds.
}\label{tab:distribution}\small
\adjustbox{max width=\linewidth}{%
\begin{tabular}{l|ccc}
\toprule
Distribution
    & \( L^2 \) distance (\( \downarrow \))
    & Clean accuracy (\%)
    & Attack success rate (\%) \\
\midrule
Uniform (\( s=0.70 \))
    & \( 1.50 \pm 0.05 \)
    & \( 94.51 \pm 0.32 \)
    & \( 92.66 \pm 0.52 \) \\
Uniform (\( s=0.75 \))
    & \( 1.61 \pm 0.07 \)
    & \( 94.22 \pm 0.12 \)
    & \( 93.74 \pm 0.66 \) \\
Beta (\( \beta = 2 \))
    & \( 1.67 \pm 0.07 \)
    & \( 94.29 \pm 0.14 \)
    & \( 97.25 \pm 0.63 \) \\
Uniform (\( s = 0.8\))
    & \( 1.77 \pm 0.09 \)
    & \( 94.21 \pm 0.22 \)
    & \( 95.51 \pm 1.04 \) \\
Gaussian (\( \sigma = 0.5 \))
    & \( 1.84 \pm 0.06 \)
    & \( 94.73 \pm 0.09 \)
    & \( 91.24 \pm 2.13 \) \\
Beta (\( \beta = 1 \))
    & \( 2.04 \pm 0.12 \)
    & \( 94.41 \pm 0.08 \)
    & \( 98.80 \pm 0.07 \) \\
Gaussian (\( \sigma = 0.75 \))
    & \( 2.74 \pm 0.11 \)
    & \( 94.13 \pm 0.14 \)
    & \( 95.17 \pm 0.76 \) \\
\bottomrule
\end{tabular}}
\end{table}

\subsection{Defense Experiments}

\begin{figure}[ht]
    \centering
    \fpplot{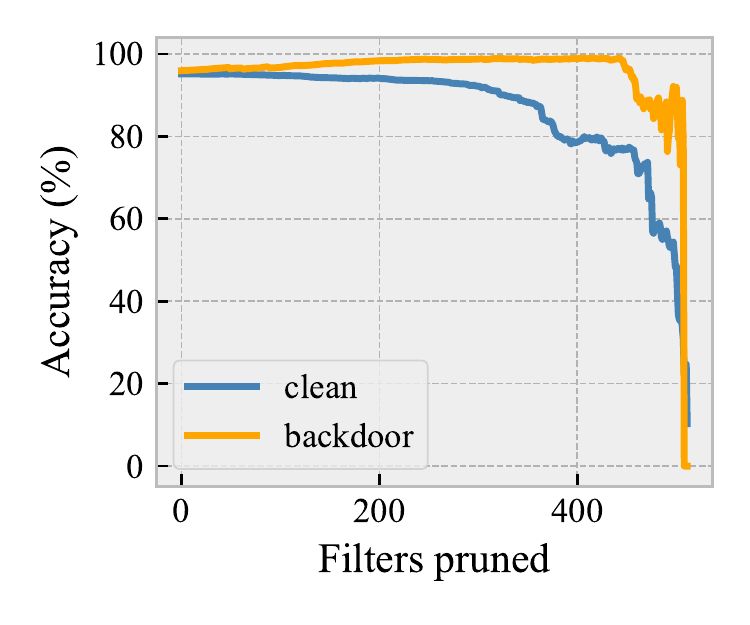}{Fine-pruning on \cifarx{}.}
    \fpplot{celeba}{Fine-pruning on \celeba{}.} \\
    \stripplot{celeba}{STRIP on \celeba{}.}
    \stripplot{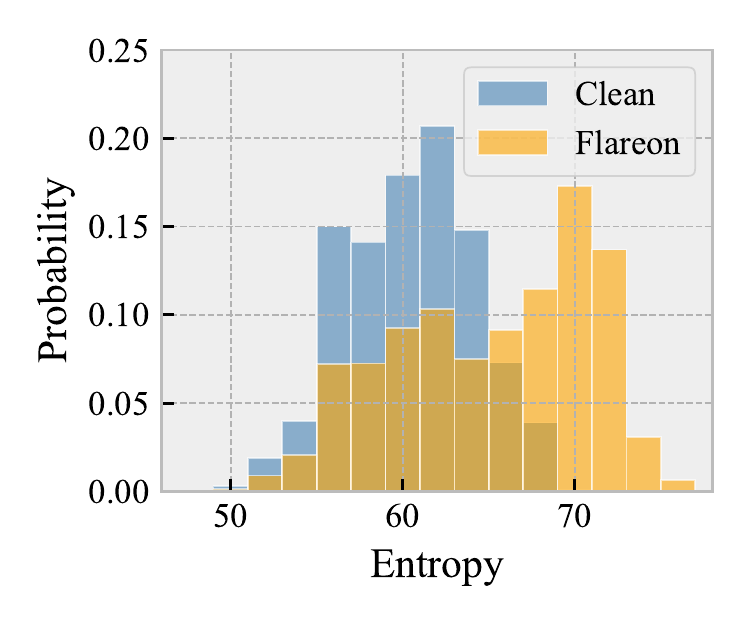}{STRIP on \tin{}.}
    \ncplot{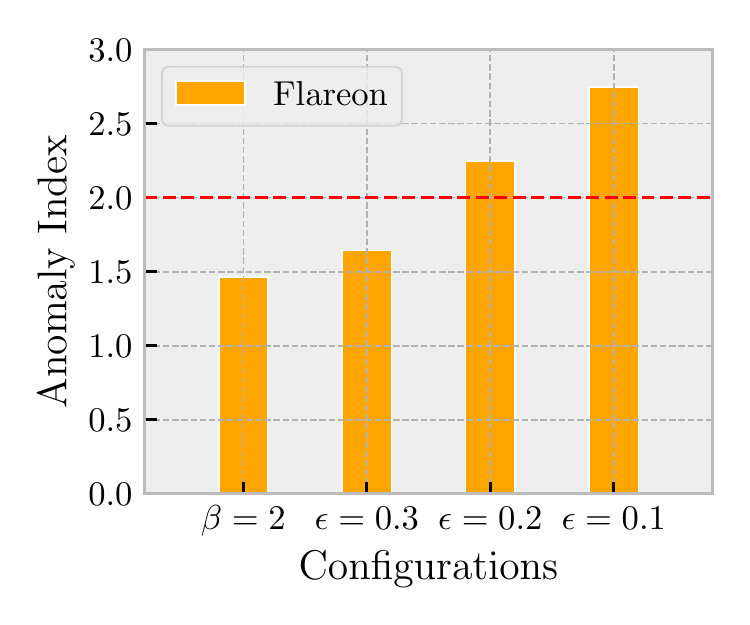}{NC on learned triggers.}
    \caption{%
        (\subref{fig:fp:cifar},
         \subref{fig:fp:celeba})
        Fine-pruning for the \cifarx{}
        and \celeba{} models.
        (\subref{fig:strip:celeba},
         \subref{fig:strip:tin})
        STRIP defenses on \celeba{}
        and \tin{} models.
        (\subref{fig:nc:configs})
        Smaller perturbations
        are easier to detect
        for neural cleanse.
    }\label{fig:defense:extra}
\end{figure}

\begin{figure}[hbt]
    \centering
    \includegraphics[
        scale=0.38, trim=0 30pt 0 0
    ]{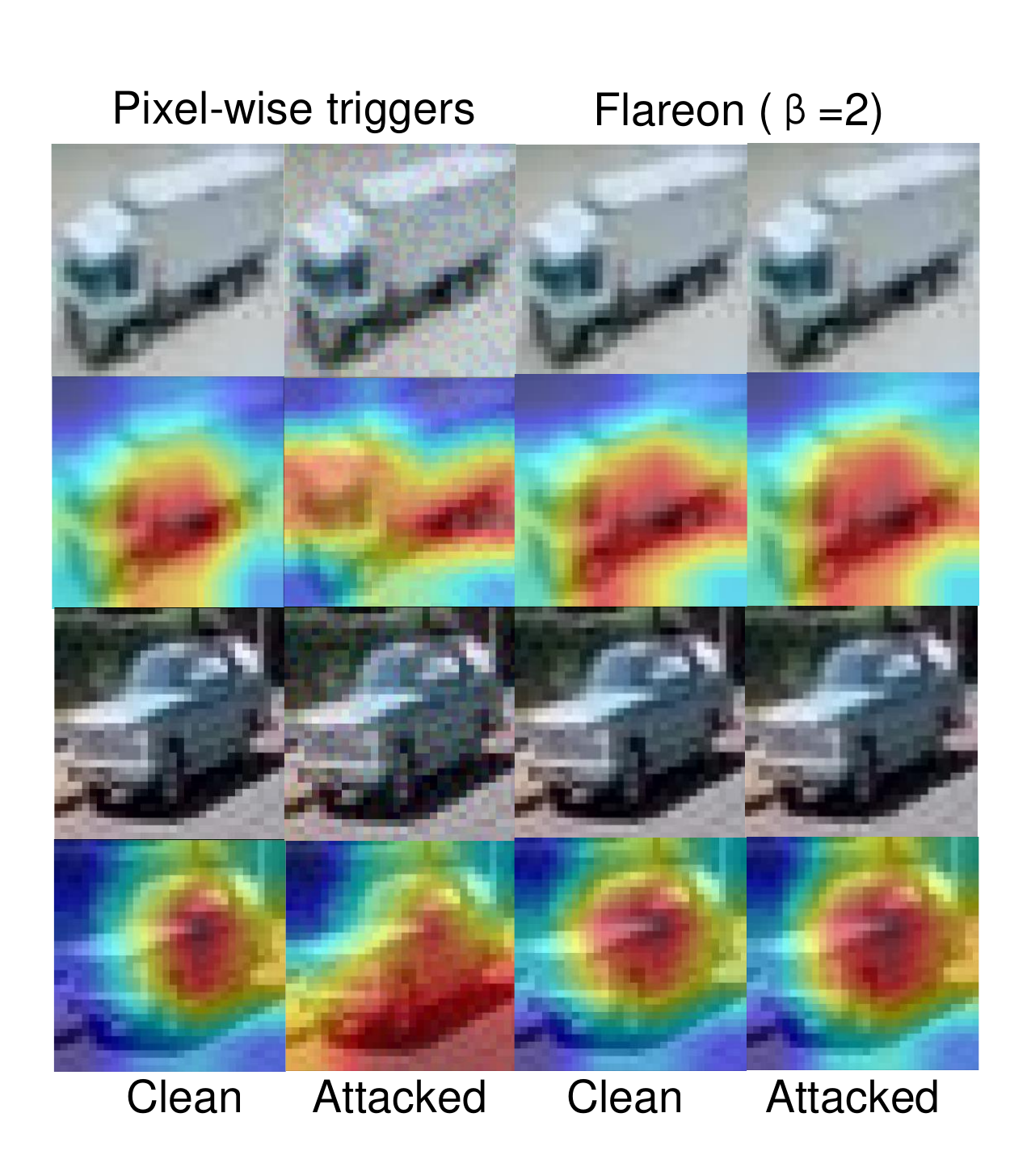}
    \caption{%
        Grad-CAM heat maps and perturbed images
        comparisons between \Method{} and pixel-wise triggers.}
    \label{fig:gradcam}
\end{figure}

\Cref{fig:defense:extra}
provides additional defense results.
Visualization tools
such as Grad-CAM~\citep{selvaraju2017grad}
are helpful
in providing visual explanations
of neural networks.
Following~\cite{nguyen2020wanet},
we also evaluate the behavior
of backdoored models against such tools.
Pixel-wise triggers as used in~\Cref{tab:ablation}
are easily exposed due
to its fixed trigger pattern
(\Cref{fig:gradcam}).

\rebuttal{%
To demonstrate the reliability
of \Method{} under randomized smoothing,
we apply \cite{wang20randsmooth} on \Method{}
with different trigger proportions \( \rho \),
as shown in the \Cref{tab:smooth}.
In addition,
we follow the setup of RAB~\citep{weber2020rab},
an ensemble-based randomized smoothing defense,
and use the official implementation
for empirical robustness evaluation,
which sets the number of sampled
noise vectors to \( N = 1000 \),
and samples the smoothing noise
from the Gaussian distribution
\( \mathcal{N}\parens{0, 0.2} \)
on \cifarx{}.
For fairness,
we use the same CNN model
and evaluation methodology
in RAB\@.
The experimental results are in~\Cref{tab:rab}.
\Method{} enjoys great success
under smoothing-based defenses.}
\begin{table}[ht]
\centering\caption{%
    Evaluation of randomized smoothing on \Method{}.
}\label{tab:smooth}\small
\begin{tabular}{llccccc}
\toprule
Model
    & \multicolumn{1}{r}{\( \rho = {} \)}
    & 50 & 60 & 70 & 80 & 90 \\
\midrule
\multirow{2}{*}{\cifarx{}}
    & Clean accuracy (\%)
    & 92.24 & 87.82 & 85.38 & 76.37 & 63.72 \\
    & Attack success rate (\%)
    & 97.33 & 96.70 & 98.10 & 99.42 & 99.16 \\
\bottomrule
\end{tabular}
\vspace{15pt}
\caption{%
    Evaluation of RAB on \Method{}.
    ``Vanilla'' denotes training without RAB\@.
    Following~\cite{weber2020rab}
    for evaluation,
    the empirical robust accuracy
    reports the proportion
    of malicious inputs
    that not only attacks the vanilla model successfully,
    but also tricks RAB\@.
}\label{tab:rab}\small
\begin{tabular}{lcc|ccccc}
\toprule
Model
    & \multicolumn{2}{c|}{Benign Accuracy (\%)}
    & \multicolumn{5}{c}{%
        Empirical Robust Accuracy under \Method{} (\%)
    } \\
\midrule
\multirow{2}{*}{\cifarx{}}
    & Vanilla
    & RAB
    & Vanilla
    & \( \rho = 50\% \) & \( \rho = 60\% \)
    & \( \rho = 70\% \) & \( \rho = 80\% \) \\
{}
    & 61.71 & 58.74 & 0 & 9.71 & 8.15 & 6.45 & 3.82 \\
\bottomrule
\end{tabular}
\end{table}

\FloatBarrier%
\subsection{Discussion and Results of \cite{turner2019label}}

\rebuttal{%
Label-consistent backdoor attacks (LC)~\cite{turner2019label}
encourages the model to learn backdoors
by generating poisoned examples without altering their labels.
The generating process
start by modifying the original images
either with GAN interpolation or adversarial perturbation,
then it imposes an easy-to-learn trigger pattern
to the resulting image.
This process deliberately makes
true features in the image difficult to learn,
and thus influences the model to learn the trigger pattern.
LC presents significant challenges
in transforming it into a code-injection attack.
The reasons are as follows:
\begin{enumerate}
    \item
    The triggers are clearly visible
    to human (\Cref{fig:trigger_compare}).

    \item
    GAN usage assumes prior knowledge
    of the data,
    whereas \Method{} is data-agnostic.

    \item Synthesizing
    GAN-interpolated examples
    or PGD-100 adversarial examples
    requires expensive pre-computation before training.

    \item
    Even if they are directly deployed as code-injection attacks,
    run-time profiling inspections,
    \eg{}, with PyTorch profiler
    will reveal both approaches
    contain erroneous unwanted computations.
    In contrast,
    \Method{} disguises its simple operations
    as useful data augmentation,
    and is thus a lot more stealthy in this regard.

    \item
    Because of the constant triggers
    and harmful alterations to the original images,
    We show that LC
    is unlikely to be effective against NC (\Cref{fig:nc:lc}),
    and they are also impactful on clean accuracies (\Cref{tab:lc:acc}).
\end{enumerate}
Furthermore,
\Method{} introduces \anytoany{} backdoors
with clean-label training,
whereas LC limits itself to single-targeted attacks.

\begin{figure}[ht]
    \centering
    \includegraphics[scale=0.8]{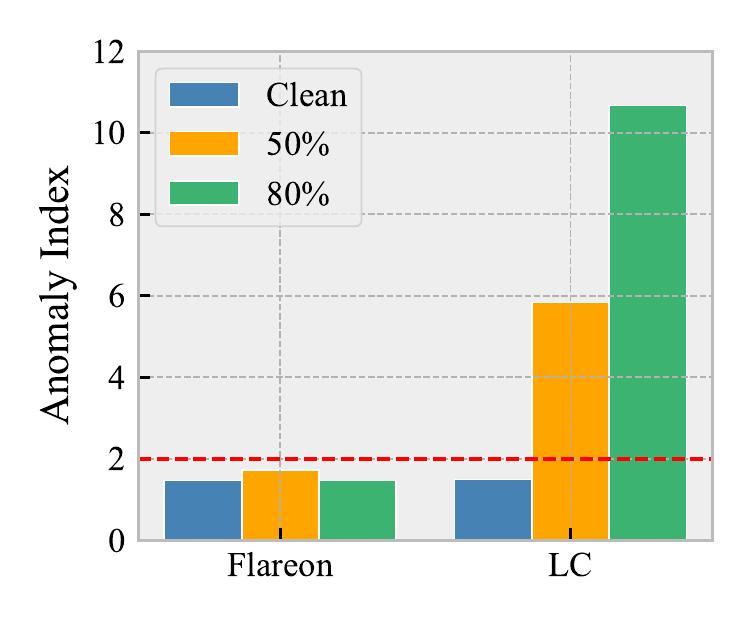}
    \caption{%
        Comparing NC on \Method{} and LC
        under different trigger proportions
        \( \rho \in \braces{50\%, 80\%} \)\@.
        Note that \Method{} attacks all classes,
        whereas LC alters images
        from the first class (``airplane'') only.
    }\label{fig:nc:lc}
\end{figure}
\begin{table}[ht]
\centering\caption{
    Comparing LC~\citep{turner2019label} with PGD-100
    and \Method{}
    on \cifarx{}
    in terms of clean accuracies.
    We remind that \( \beta \) is used
    for trigger initialization
    and larger values indicate stealthier triggers.
    Here, ``\( y \to t \)''
    means single-targeted attack.
    To compare with LC,
    we provide results
    that restrict \Method{}'s capability
    to single-target poisoning only,
    which translates to poisoning \( \rho / 10 \)
    of all training examples per mini-batch
    on \cifarx{}.
}\label{tab:lc:acc}
\adjustbox{max width=\linewidth}{%
\begin{tabular}{{ldddd|dd}}
\toprule
{}
    & \multicolumn{2}{c}{LC}
    & \multicolumn{4}{c}{\Method{}} \\
{}
    & \multicolumn{2}{c}{\( y \to t \), PGD-100}
    & \multicolumn{2}{c|}{\( y \to t \), \( \beta = 1 \)}
    & \multicolumn{2}{c}{\anytoany{}, \( \beta = 2 \)} \\
\midrule
Baseline accuracy without attack (\%)
    & \multicolumn{2}{c}{92.53}
    & \multicolumn{2}{c|}{96.14}
    & \multicolumn{2}{c}{96.14} \\
\cmidrule{1-7}
Average poisoned samples per batch
    & \tcenter{5\%} & \multicolumn{1}{c}{8\%}
    & \tcenter{5\%} & \multicolumn{1}{c|}{8\%}
    & \tcenter{50\%} & \tcenter{80\%} \\
\cmidrule{1-7}
Clean accuracy (\%)
    & 89.61 & 89.30
    & \tbnum{95.70} & 94.41
    & 94.22 & 94.43 \\
\( \Delta \) Clean accuracy (\%)
    & -2.92 & -3.23
    & \tbnum{-0.44} & -1.73
    & -1.92 & -1.71 \\
Attack success rate (\%)
    & 81.47 & 96.00
    & 85.32 & 98.60
    & 93.14 & 97.78   \\
\bottomrule
\end{tabular}}
\end{table}
}


\end{document}